\documentclass[12pt,preprint]{aastex}
\shorttitle{The Star Formation Law at Low Density}
\shortauthors{Wyder et al.}

\begin{document}

\title{The Star Formation Law at Low Surface Density}

\author{Ted K. Wyder\altaffilmark{1},
D. Christopher Martin\altaffilmark{1},
Tom A. Barlow\altaffilmark{1},
Karl Foster\altaffilmark{1},
Peter G. Friedman\altaffilmark{1},
Patrick Morrissey\altaffilmark{1},
Susan G. Neff\altaffilmark{2},
James D. Neill\altaffilmark{1}, 
David Schiminovich\altaffilmark{3},
Mark Seibert\altaffilmark{4},
Luciana  Bianchi\altaffilmark{5},
Jos\'e Donas\altaffilmark{6},
Timothy M. Heckman\altaffilmark{7},
Young-Wook Lee\altaffilmark{8},
Barry F. Madore\altaffilmark{4},
Bruno Milliard\altaffilmark{6},
R. Michael Rich\altaffilmark{9},
Alex S. Szalay\altaffilmark{7}, \\
and Sukyoung K. Yi\altaffilmark{8}
}

\altaffiltext{1}{California Institute of Technology, MC 405-47, 1200 E. California Blvd., Pasadena, CA 91125}
\altaffiltext{2}{Laboratory for Astronomy and Solar Physics, NASA  Goddard Space Flight Center, Greenbelt, MD, 20771}
\altaffiltext{3}{Department of Astronomy, Columbia University, New  York, NY 10027}
\altaffiltext{4}{Observatories of the Carnegie Institution of Washington, 813 Santa Barbara St., Pasadena, CA 91101}
\altaffiltext{5}{Center for Astrophysical Sciences, The Johns Hopkins University, 3400 N. Charles St., Baltimore, MD, 21218}
\altaffiltext{6}{Laboratoire d'Astrophysique de Marseille, BP 8, Traverse du Siphon, 13376 Marseille Cedex 12, France}
\altaffiltext{7}{Department of Physics and Astronomy, The Johns Hopkins University, Homewood Campus, Baltimore, MD 21218}
\altaffiltext{8}{Center for Space Astrophysics, Yonsei University, Seoul 120-749, Korea}
\altaffiltext{9}{Department of Physics and Astronomy, University of California, Los Angeles, CA 90095}

\begin{abstract}
We investigate the nature of the star formation law at low gas surface densities using a sample of 19 low surface brightness (LSB) galaxies with existing \ion{H}{1} maps in the literature, UV imaging from the {\it Galaxy Evolution Explorer} satellite, and optical images from the Sloan Digital Sky Survey. All of the LSB galaxies have $(NUV-r)$ colors similar to those for higher surface brightness star-forming galaxies of similar luminosity indicating that their average star formation histories are not very different. Based upon four LSB galaxies with both UV and FIR data, we find FIR/UV ratios significantly less than one, implying low amounts of internal UV extinction in LSB galaxies. We use the UV images and \ion{H}{1} maps to measure the star formation rate and hydrogen gas surface density within the same region for all of the galaxies. The LSB galaxy star formation rate surface densities lie below the extrapolation of the power law fit to the star formation rate surface density as a function of the total gas density for higher surface brightness galaxies. Although there is more scatter, the LSB galaxies also lie below a second version of the star formation law in which the star formation rate surface density is correlated with the gas density divided by the orbital time in the disk. The downturn seen in both star formation laws is consistent with theoretical models that predict lower star formation efficiencies in LSB galaxies due to the declining molecular fraction with decreasing density. 
\end{abstract}

\section{Introduction}

Understanding what regulates the formation of stars in galaxies is one of the most important aspects of understanding the evolution of galaxies. In particular the relationship between the amount of gas in galaxies and their star formation rates is an important ingredient in galaxy simulations that include the evolution of the baryons in addition to the dark matter. The lack of a theory of star formation in galaxies means that simulations must rely upon empirical correlations to determine the parameters affecting star formation in their models \citep[e.g.,][]{springel2003}.

Empirically, it has been shown that the star formation rates and gas content in galaxies are related via the Schmidt-Kennicutt Law \citep{kennicutt1989, kennicutt1998a} given by
\begin{equation}
\Sigma_{SFR}  = A \Sigma_{gas}^N,
\label{sflaw1_eqn}
\end{equation}
where $\Sigma_{SFR}$ and $\Sigma_{gas}$ are the star formation rate (SFR) and total gas surface densities, respectively. Based upon a sample of spiral, starburst, and luminous infrared galaxies spanning gas surface densities in the range $10-10^5$ $M_{\odot}$ pc$^{-2}$, \citet{kennicutt1998a} found a best-fit exponent of $N = 1.4 \pm 0.15$. However, an alternative expression for the star formation law that describes the observations equally well is given by
\begin{equation}
\Sigma_{SFR} = B \frac{\Sigma_{gas}}{\tau_{dyn}},
\label{sflaw2_eqn}
\end{equation}
where $\tau_{dyn}$ is a characteristic dynamical time usually defined as the orbital time of the star-forming disk \citep{kennicutt1998a}. 

More recent investigations of the star formation law using resolved data in nearby galaxies have shown a better correlation between $\Sigma_{SFR}$ and the molecular gas surface density rather than the total gas surface density although different studies have derived different values for the exponent $N$ in equation (\ref{sflaw1_eqn}). For example, based upon resolved data for the galaxy M51, \citet{kennicutt2007} found a good correlation between $\Sigma_{SFR}$, as traced by H$\alpha$ and $24\micron$ data, and the molecular gas surface density and very little correlation with the atomic gas surface density. The best-fitting power law as a function of $\Sigma_{H2}$ alone in M51 has an exponent in the range $N=1.4-1.6$, depending upon the scale upon with the SFR and molecular gas densities are measured. Similar to these results, \citet{heyer2004} found a good correlation between the SFR surface density, as traced by the FIR emission, and the molecular gas surface density in M33 with an exponent of $N=1.4$. On the other hand, if the unobscured star formation traced by the UV is included in the estimate of the SFR surface density in M33, the correlation with the molecular gas surface density becomes somewhat shallower \citep[i.e. smaller $N$;][]{gardan2007}. \citet{bigiel2008} and \citet{leroy2008} found similar results for a sample of spiral and irregular galaxies with resolved gas and SFR density maps except with a best-fit exponent of $N=1.0\pm0.2$. 

In most galaxies there appears to be a threshold in the gas surface density in the range $3-10$ $M_{\odot}$ pc$^{-2}$ below which very little star formation is observed, at least as traced by the H$\alpha$ emission line \citep{kennicutt1989, kennicutt1998a, martin2001}. Using azimuthally averaged radial gas and star formation rate profiles, \citet{kennicutt1989, kennicutt1998a} found that in the vicinity of the threshold density, the relation between SFR density and gas density is steeper than the 1.4 power law of equation (\ref{sflaw1_eqn}). This increase in the slope of the star formation law has also been observed in the outer regions of the high surface brightness spiral galaxies NGC 628 and NGC 7331 \citep{lelievre2000, thilker2007}. The break in the star formation law in both galaxies coincides with the radius where the gas transitions from being predominantly molecular to mostly atomic. On the other hand, based upon FIR and UV radial profiles in a sample of 43 galaxies, \citet{boissier2007} found that the star formation law does not show a sharp down turn or threshold at low density although the scatter is large. Thus, it is important to obtain additional measurements of star formation at low densities to better constrain the nature of both forms of the star formation law in the regime of the gas density threshold.

The value of the star formation density threshold varies from galaxy to galaxy and is usually explained in terms of the Toomre criterion which determines the local gas density threshold below which the gas is stable to collapse and star formation \citep{martin2001}. There are however other explanations. \citet{schaye2004} has argued that the observations can be equally well explained by a gas density threshold determined by the radius at which a cold phase in the ISM can form.  On the other hand, \citet{blitz2004} have contended that $\Sigma_{SFR}$ in galaxies is more closely related to the amount of molecular gas which is in turn determined by the local hydrostatic pressure in the interstellar medium. 

There have been various theoretical explanations proposed to explain the star formation laws as expressed in equations (\ref{sflaw1_eqn}) and (\ref{sflaw2_eqn}). In particular, \citet{krumholz2005} have proposed a theory in which star formation in galaxies occurs primarily in regions of supersonically turbulent molecular clouds where the local gravitational potential energy is able to overcome the turbulence. They provide analytical predictions for the star formation rates in galaxies that are consistent with both forms of the star formation law in equations (\ref{sflaw1_eqn}) and (\ref{sflaw2_eqn}). This model predicts a downturn in both forms of the star formation law at low total (atomic plus molecular) gas density due to the declining molecular fraction with decreasing density. Based upon a set of hydrodynamical simulations of disk galaxies \citet{robertson2008} also predict a down-turn in the star formation rates at low density. 

In order to better understand the star formation law and to test the predictions of the various theoretical models, it is important to measure the star formation law over as large a range in gas density as possible. In this paper, we extend the star formation law to gas densities below 10 $M_{\odot}$ pc$^{-2}$ using a sample of low surface brightness (LSB) galaxies with resolved \ion{H}{1} data from the literature and with UV data from the {\it Galaxy Evolution Explorer (GALEX)} satellite. The plan of this paper is as follows. In \S2 we describe the sample of LSB galaxies as well as the UV and optical images of them. In \S3 we describe our results on extending the star formation law to low density. Finally, we conclude in \S4.

\section{Data}

In order to test the nature of the star formation law at low surface density, we compiled a sample of 19 LSB galaxies with existing resolved \ion{H}{1} data in the literature and UV imaging from the {\it GALEX} satellite. {\it GALEX} is a 50cm diameter UV telescope in low Earth orbit observing the sky simultaneously in two bands, a $FUV$ band centered at 1450\AA~and a $NUV$ band centered at 2300\AA~\citep{morrissey2005,morrissey2007,martin2005}. {\it GALEX} is conducting surveys of large portions of the sky as well as targeted observations of many nearby galaxies. The UV data presented here is a part of the {\it GALEX} GR4 data release made publicly available through the Multi-Mission Archive at the Space Telescope Science Institute (MAST).\footnote{The GR4 data release can be accessed via the web address http://galex.stsci.edu/GR4/.}

The largest portion of the sample (14) comes from \citet{deblok1996}. These galaxies were discovered originally by \citet{schombert1992} from a visual search of the Second Palomar Sky Survey blue plates for LSB galaxies with diameters greater than $30\arcsec$. From this catalog \citet{deblok1996} selected a sample of LSB galaxies with \ion{H}{1} masses in the range $8.5 <  \log{M_{HI}/M_{\sun}}  <  9.5$ and redshifts between 3000 and 8000 km s$^{-1}$ to observe with either the Very Large Array or the Westerbork Synthesis Radio Telescope. Of the 19 galaxies in their paper, 14 have observations with {\it GALEX} and all are detected in the UV. We have supplemented these galaxies with two additional LSB galaxies, UGC 5750 and UGC 5999, that have both {\it GALEX} imaging and resolved \ion{H}{1} maps from \citet{vanderhulst1993}. Finally, we also included in our sample three giant LSB galaxies (LSBC F568-06, Malin 1, and UGC 6614) with {\it GALEX} observations and \ion{H}{1} maps from \citet{pickering1997}. 

For the galaxies from the \citet{pickering1997} and \citet{vanderhulst1993} papers, the luminosity distances were calculated from the radial velocities presented in those papers assuming a Hubble constant of $H_0 = 70$ km s$^{-1}$ Mpc$^{-1}$, $\Omega_{matter}=0.3$, and $\Omega_{\Lambda}  = 0.7$. For the galaxies from \citet{deblok1996}, we assumed the distances presented in that paper except converted to $H_0=70$ km s$^{-1}$ Mpc$^{-1}$. The positions and luminosity distances for our LSB galaxy sample are listed in Table \ref{sample}. The {\it GALEX} UV observations of the sample consist of data from the All-Sky-Imaging Survey, Nearby Galaxies Survey, and various Guest Investigator programs and the exposure times and field names for each are listed in Table \ref{uvobs}.

\begin{figure}
\includegraphics[width=6in,height=8in]{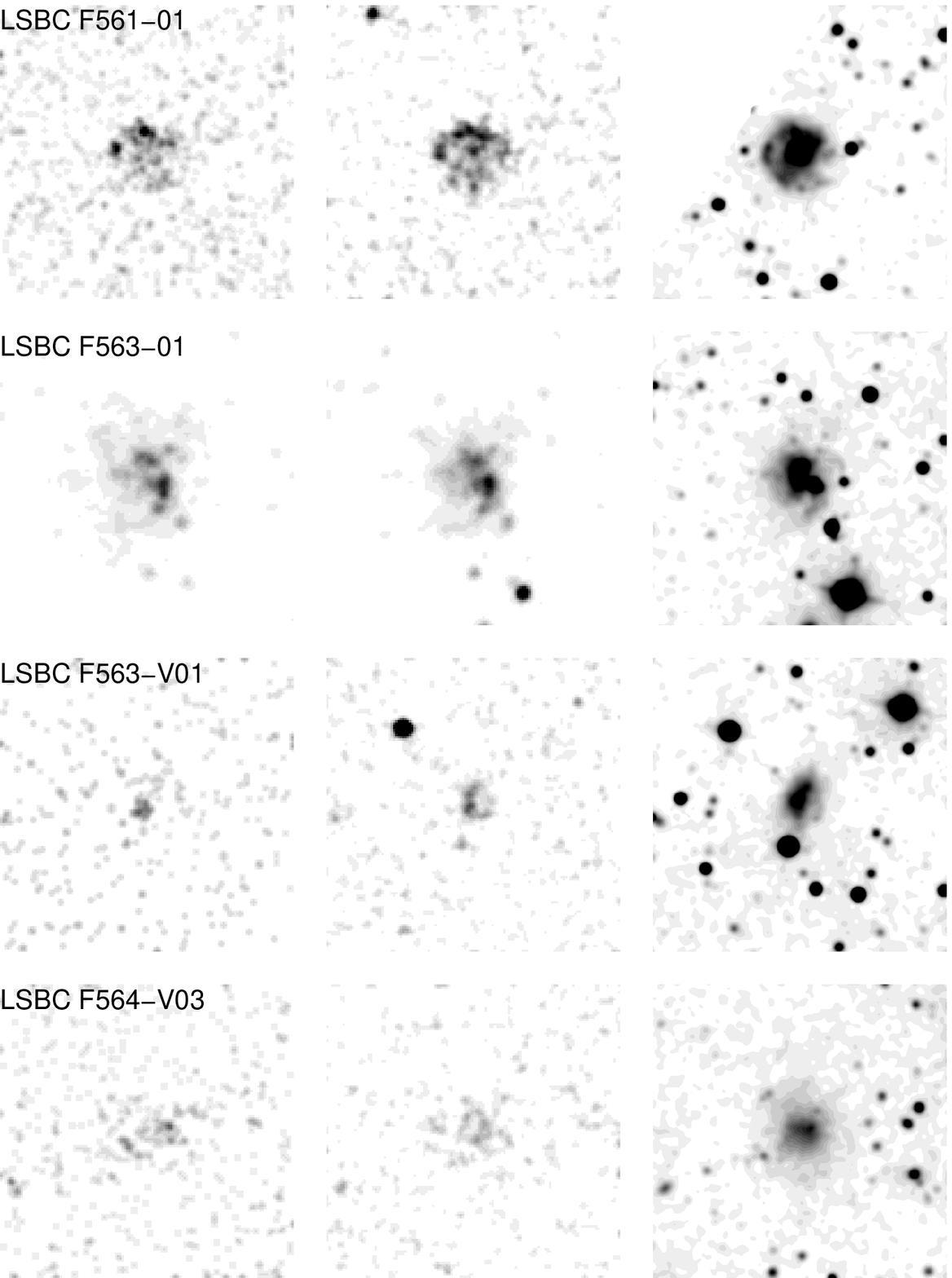}
\caption{Images of low surface brightness galaxies from \citet{deblok1996} in the {\it GALEX} $FUV$ (left) and $NUV$ (middle) and in the SDSS $r$-band (right). All the images are $3.2\arcmin \times 3.2\arcmin$ in size and have north up and east to the left. All of the images have been convolved with a FWHM=4.5$\arcsec$ Gaussian. 
\label{deblokimages1}}
\end{figure}

\begin{figure}
\includegraphics[width=6in,height=8in]{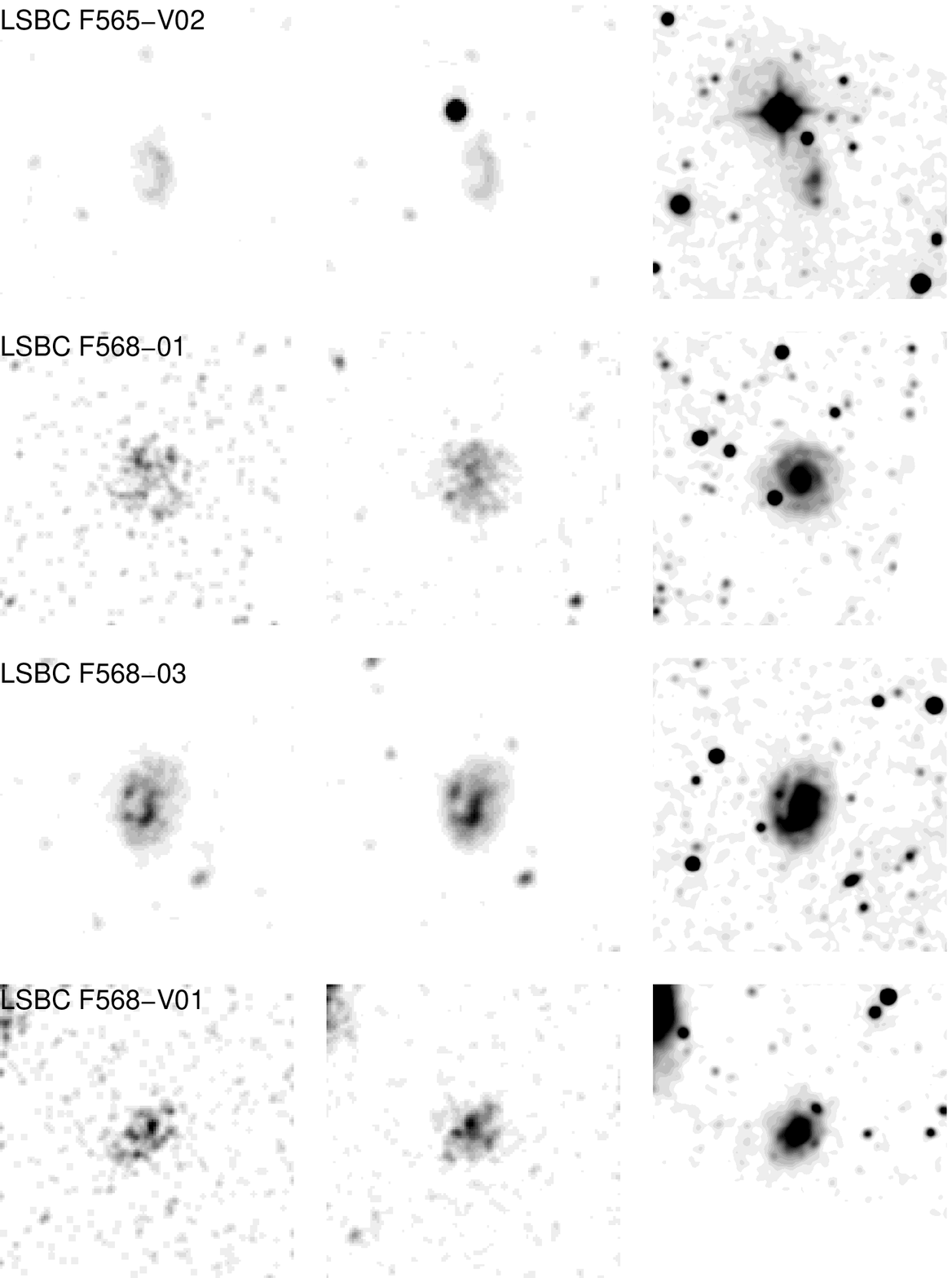}
\caption{Images of low surface brightness galaxies from \citet{deblok1996} in the {\it GALEX} $FUV$ (left) and $NUV$ (middle) and in the SDSS $r$-band (right). All the images are $3.2\arcmin \times 3.2\arcmin$ in size and have north up and east to the left. All of the images have been convolved with a FWHM=4.5$\arcsec$ Gaussian.
\label{deblokimages2}}
\end{figure}

\begin{figure}
\includegraphics[width=6in,height=8in]{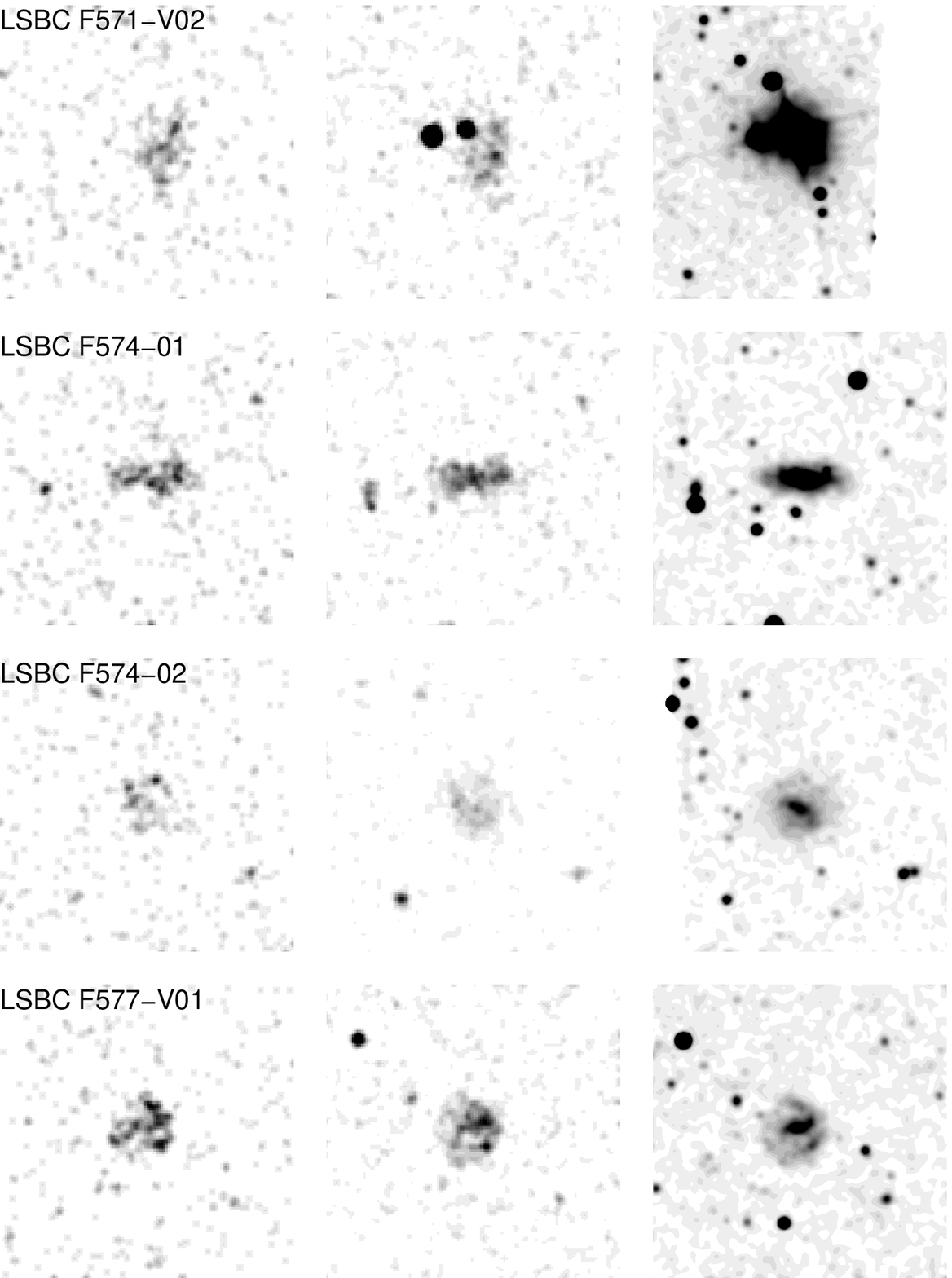}
\caption{Images of low surface brightness galaxies from \citet{deblok1996} in the {\it GALEX} $FUV$ (left) and $NUV$ (middle) and in the SDSS $r$-band (right). All the images are $3.2\arcmin \times 3.2\arcmin$ in size and have north up and east to the left. All of the images have been convolved with a FWHM=4.5$\arcsec$ Gaussian.
\label{deblokimages3}}
\end{figure}

\begin{figure}
\includegraphics[width=6in,height=4in]{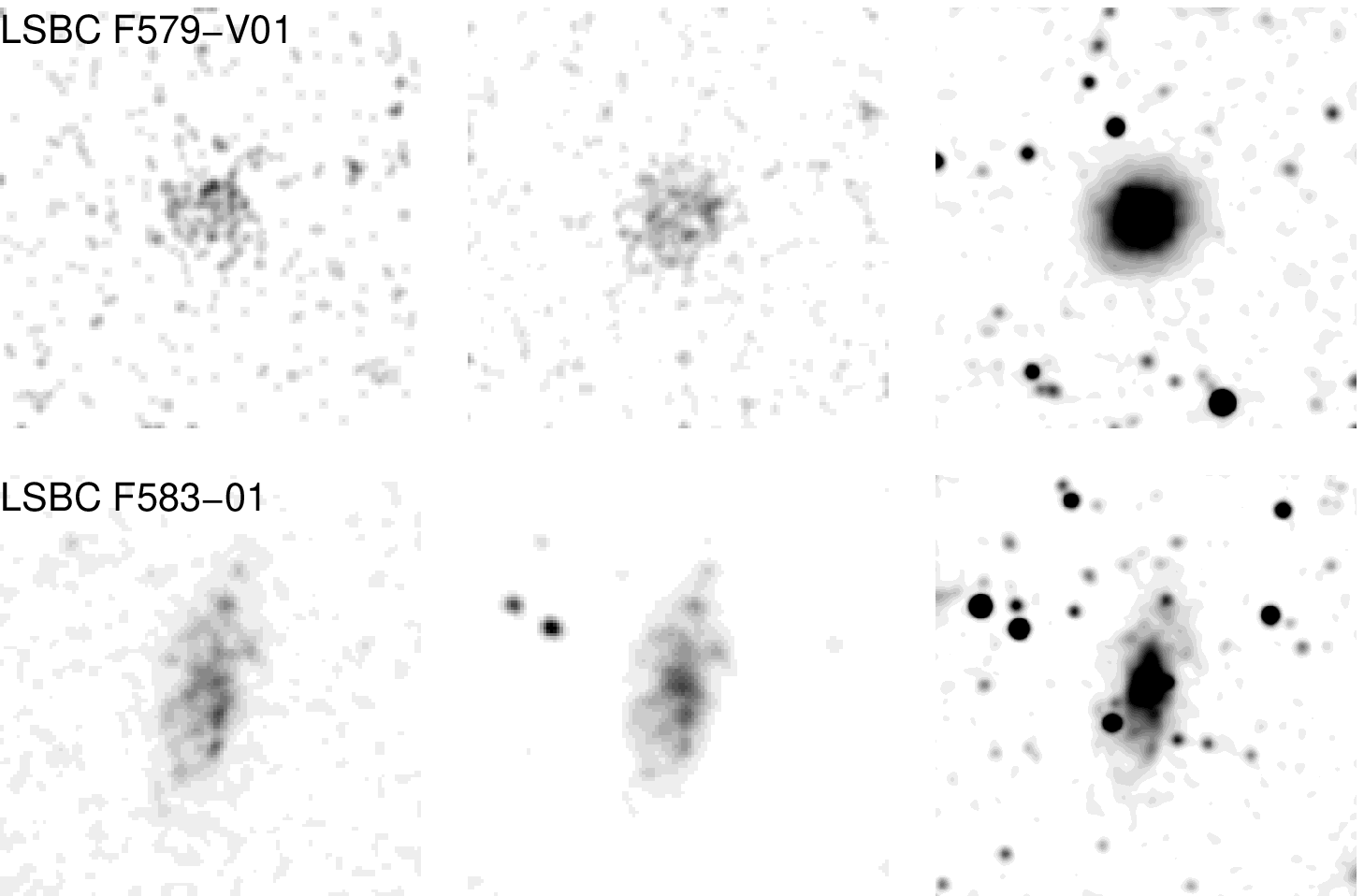}
\caption{Images of low surface brightness galaxies from \citet{deblok1996} in the {\it GALEX} $FUV$ (left) and $NUV$ (middle) and in the SDSS $r$-band (right). All the images are $3.2\arcmin \times 3.2\arcmin$ in size and have north up and east to the left. All of the images have been convolved with a FWHM=4.5$\arcsec$ Gaussian.
\label{deblokimages4}}
\end{figure}

\begin{figure}
\plotone{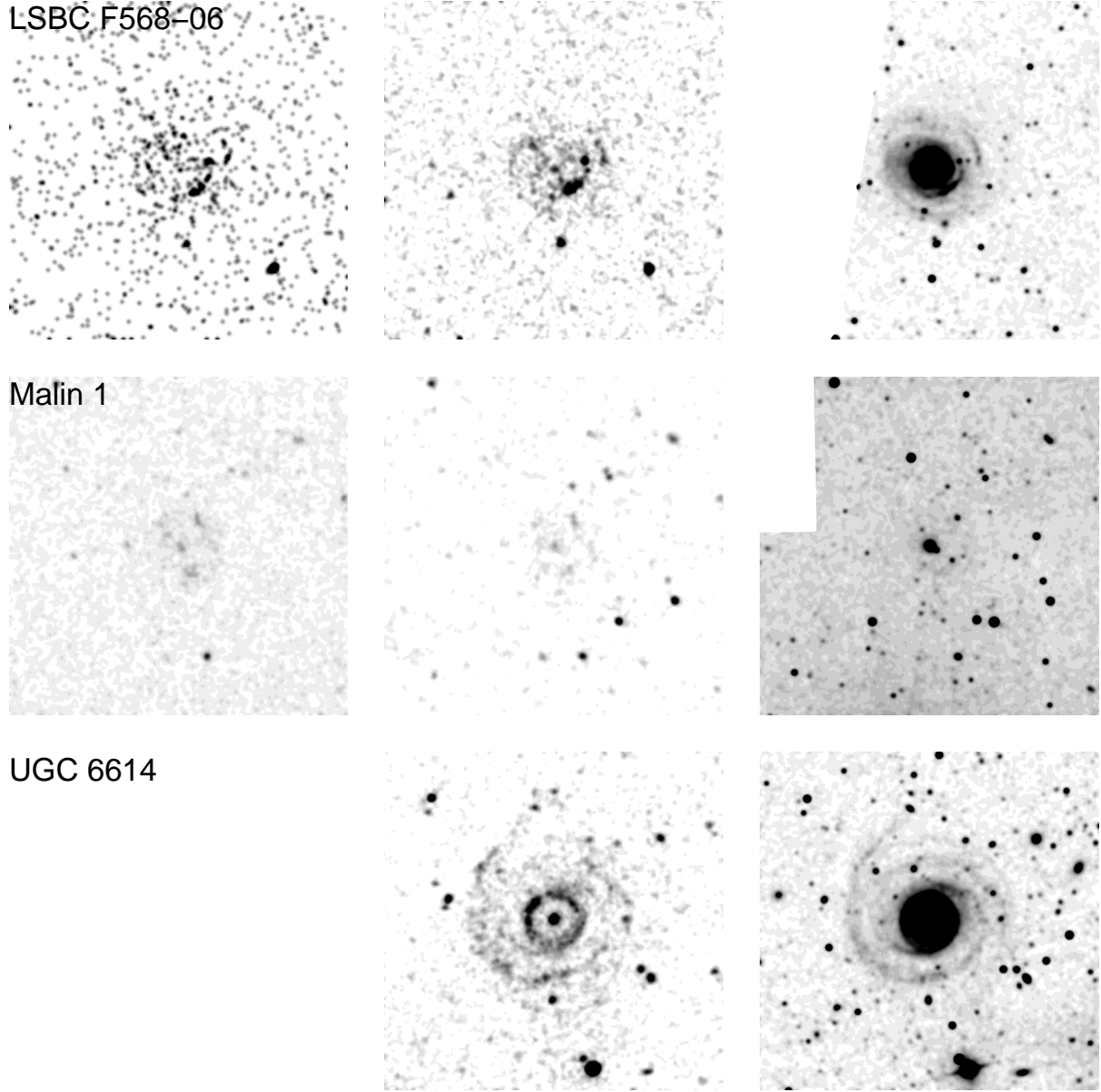}
\caption{Images of the three giant low surface brightness galaxies LSBC F568-06, Malin 1, and UGC 6614 from \citet{pickering1997} in the {\it GALEX} $FUV$ (left) and $NUV$ (middle) and in the SDSS $r$-band (right). UGC 6614 has no $FUV$ data. All the images are $6.4\arcmin \times 6.4\arcmin$ in size and have north up and east to the left. All of the images have been convolved with a FWHM=4.5$\arcsec$ Gaussian.
\label{pickering_images}}
\end{figure}

\begin{figure}
\plotone{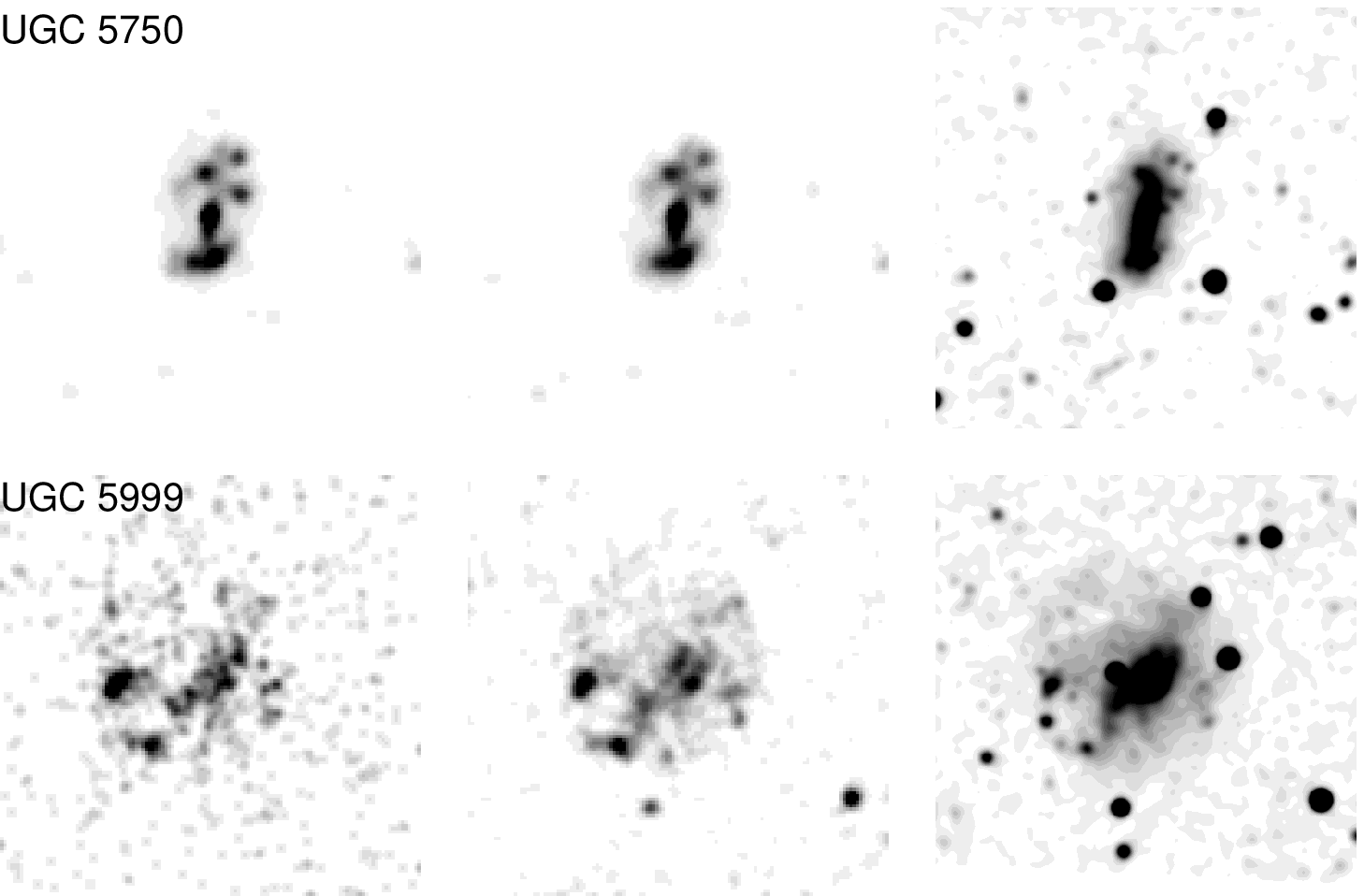}
\caption{Images of the low surface brightness galaxies UGC 5750 and UGC 5999 from \citet{vanderhulst1993} in the {\it GALEX} $FUV$ (left) and $NUV$ (middle) and in the SDSS $r$-band (right). All the images are $3.2\arcmin \times 3.2\arcmin$ in size and have north up and east to the left. All of the images have been convolved with a FWHM=4.5$\arcsec$ Gaussian.
\label{vanderhulst_images}}
\end{figure}

In addition to the UV data, all of our LSB galaxies lie within the region of sky covered by the Sloan Digital Sky Survey (SDSS). For each galaxy, we have extracted the $r$ band images from the SDSS Data Release 6. The $FUV$, $NUV$, and $r$-band images of each of the galaxies from \citet{deblok1996} are shown in Figures \ref{deblokimages1}$-$\ref{deblokimages4}. Images of the galaxies from \citet{pickering1997} and \citet{vanderhulst1993} are shown in Figures \ref{pickering_images} and \ref{vanderhulst_images}, respectively. In all the figures, the UV and optical images have been convolved with a Gaussian with FWHM of $4.5\arcsec$ in order to better view some of the fainter emission and to match the resolution of the GALEX images. Many of the galaxies have a higher surface brightness redder inner region surrounded by a more diffuse and bluer outer region. Some of the galaxies exhibit spiral arms in the optical. With the exception of the giant LSB galaxy UGC 6614, this spiral structure is not really apparent in the UV images.

Since these galaxies do have such low surface brightnesses, it is necessary to measure the sky background accurately. For each galaxy we determined the average sky value in a circular annulus around each LSB galaxy. To determine the inner radius of each sky annulus, we first convolved the $NUV$ image with a Gaussian with FWHM of $15\arcsec$. From this convolved image, we calculated the rms along the semi-major axis and set the inner radius of the sky annulus to be six times larger. This resulted in inner radii in the range $1\arcmin-6\arcmin$, corresponding to factors of $1.8-4.7$ times the semi-major axis of the ellipse used to measure the total flux as described below. The width of the sky annulus was always kept fixed at $1\arcmin$. The same annulus was used in both the optical and UV. In the $FUV$ and $NUV$ images, we masked off all sources detected by the standard {\it GALEX} pipeline. Then we divided the sky annulus into eight sectors in azimuth and calculated the average value with in each sector. The average and standard deviation of these eight sky values was assumed as the sky value and its uncertainty. In the $r$ images, a similar procedure was used except that sources in the sky annulus were not masked off. Instead, the distribution of values in each sky sector was iteratively clipped at $3\sigma$. The resulting sky values used for each galaxy are listed in Table \ref{photaper}. 

As we are interested in comparing the \ion{H}{1} and UV emission within the same area within each galaxy, we have adopted the same ellipse parameters used in the original papers to extract the \ion{H}{1} surface density profiles and rotation curves. The axis ratio and position angle for each galaxy is listed in Table \ref{photaper}. For both the UV and \ion{H}{1} profiles, we examined the radial profiles by eye to determine the largest radius with emission detected above the background. We set the radius $r_{max}$ used to define the total flux as the minimum of the UV and \ion{H}{1} radii. This same $r_{max}$ was used in the UV and the \ion{H}{1} for all of the galaxies and for most in the optical as well. For three  of the galaxies we set the size of the aperture in the $r$-band to be smaller than that used in the UV with semi-major axes of $50\arcsec$, $30\arcsec$, and $45\arcsec$ for LSBC F563-01, LSBC F568-V01, and Malin 1, respectively. For these galaxies, the optical emission is less extended than in the UV or \ion{H}{1} and the large apertures led to large uncertainties in the total flux. Some of the galaxies have stars or other galaxies within the apertures listed in Table \ref{photaper}. These sources  were manually masked off and their pixel values were replaced by the average radial profile for the LSB galaxy at that radius before summing the total flux within the elliptical aperture. We have used the same ellipse parameters to derive radial surface brightness profiles in each of the bands. 

The {\it GALEX} detectors are photon-counting, with zero read noise and low sky background, and thus the primary source of uncertainty in the fluxes is that due to counting statistics. All of the galaxies have a sufficient total number of counts so that the Poisson distribution can be well approximated by a Gaussian for the integrated measurements. On the other hand, the radial surface brightness profiles for some of the galaxies in the UV reach count levels in the outer regions low enough such that the full Poisson distribution must be taken into account when computing the errors. In order to estimate the uncertainty in the radial surface brightness measurements, we made use of the formulae derived using Bayesian methods for calculating the Poisson error in a flux by \citet{loredo1992}. In addition to the statistical error, we added in quadrature an absolute uncertainty in the {\it GALEX} calibration in each band of 5\% \citep{morrissey2007}. The resulting $FUV$ and $NUV$ magnitudes are listed in Table \ref{uvphot}. 

In the optical SDSS images the count rates are large enough such that the Poisson distribution can be well approximated by a Gaussian. We added in quadrature the Poisson uncertainty, the error due to the read noise, and the error in the sky value. The resulting $r$ magnitudes for our sample are given in Table \ref{sdssphot}. 

\begin{figure}
\plotone{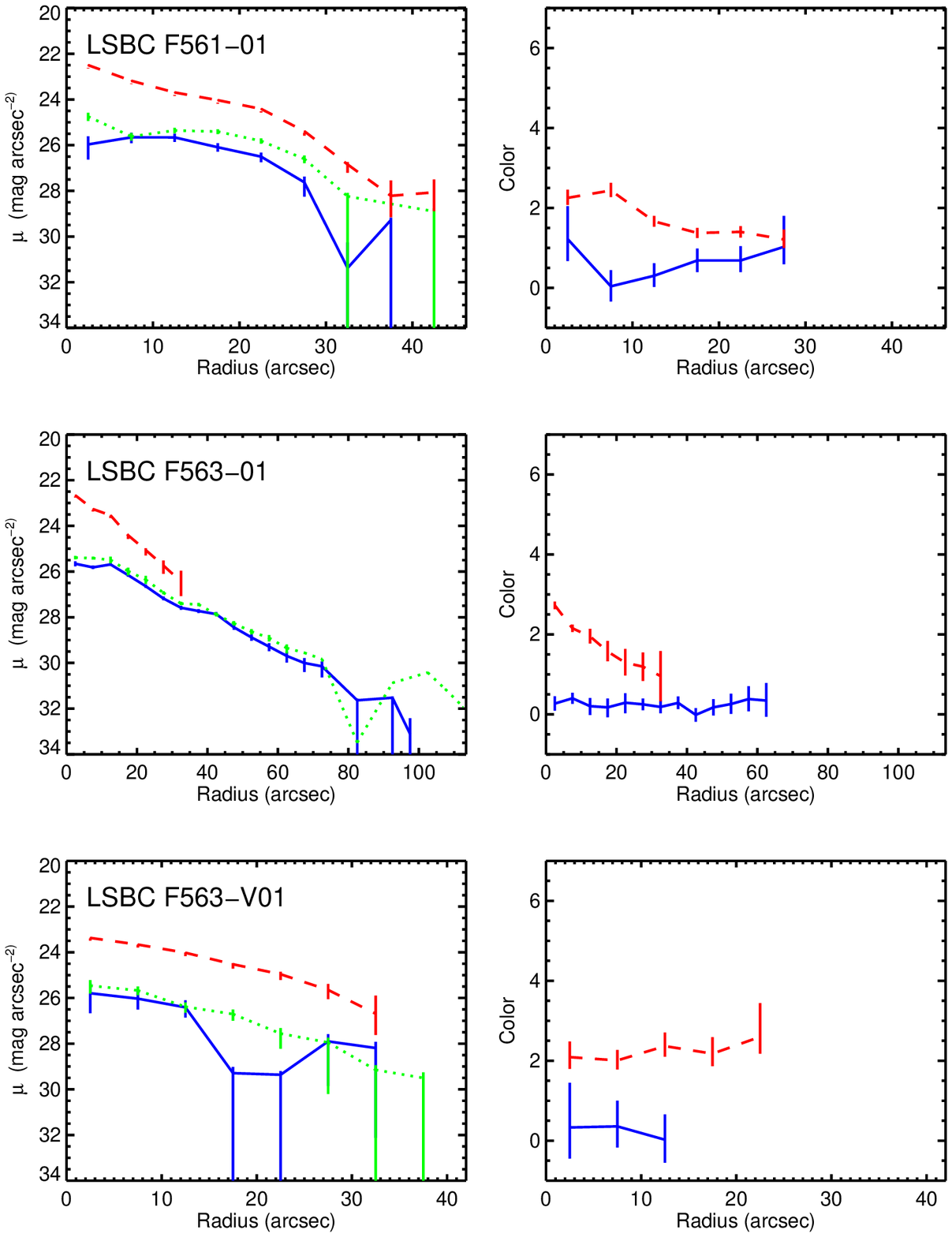}
\caption{Radial surface brightness (left) and color (right) profiles for the sample of LSB galaxies with \ion{H}{1} data from \citet{deblok1996}. The blue solid, green dotted, and dashed red curves in the left panel are the surface brightness profiles in the $FUV$, $NUV$, and $r$ bands, respectively. In the panels on the the right the $(FUV-NUV)$ color profiles are plotted as solid blue lines while the dashed red curves show the $(NUV-r)$ color. 
\label{deblok_radprof1}}
\end{figure}

\begin{figure}
\plotone{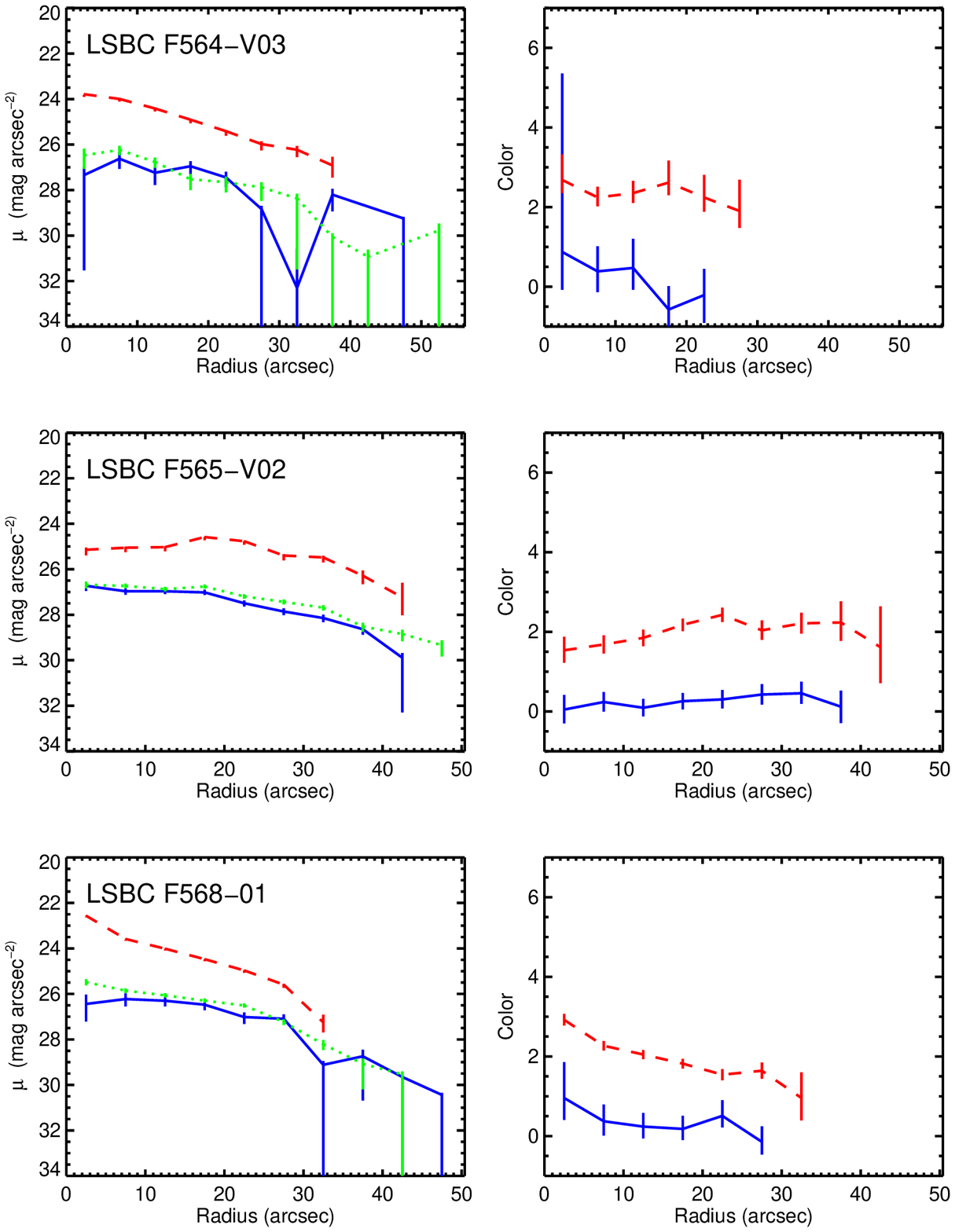}
\caption{Same as Figure \ref{deblok_radprof1}. \label{deblok_radprof2}}
\end{figure}

\begin{figure}
\plotone{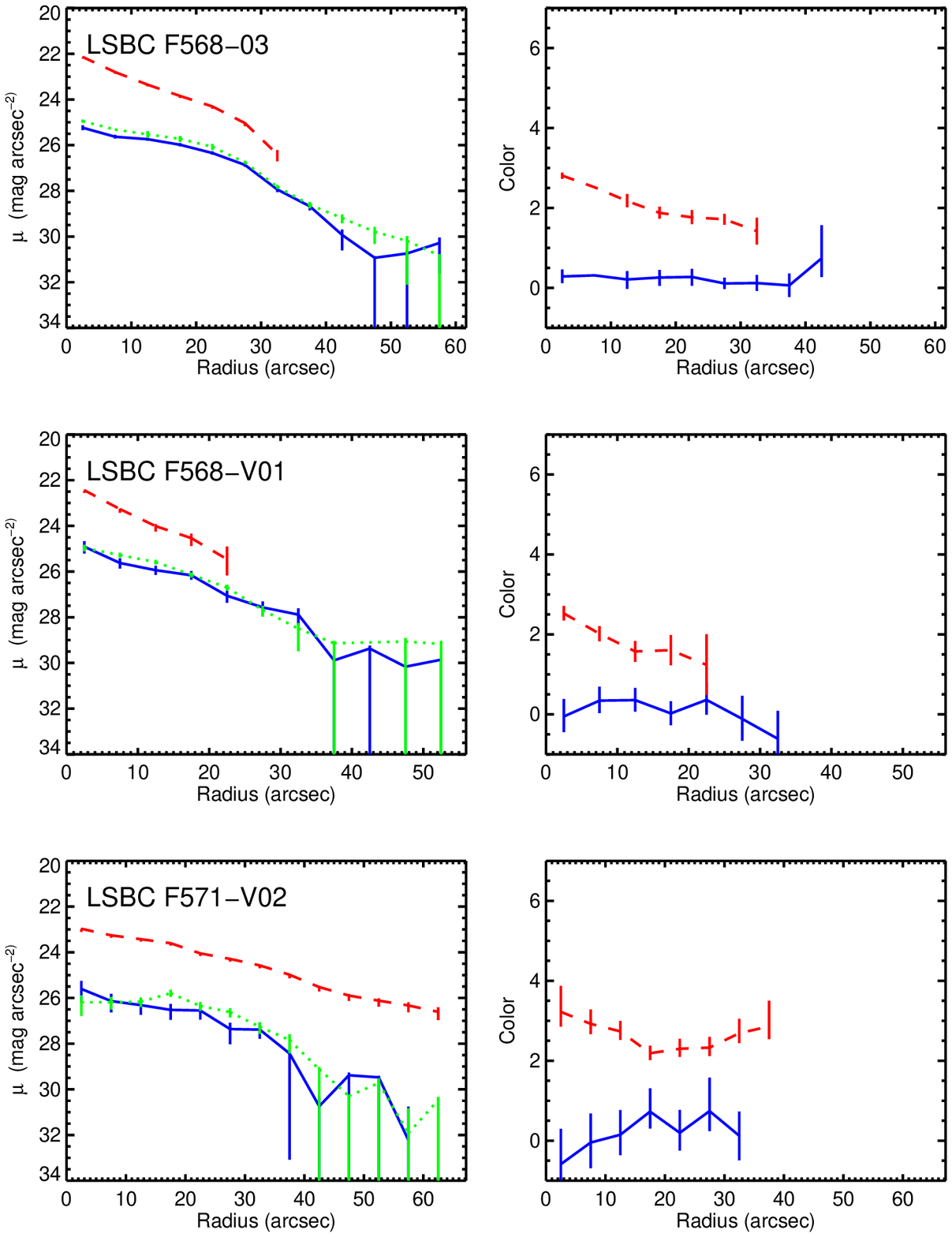}
\caption{Same as Figure \ref{deblok_radprof1}. \label{deblok_radprof3}}
\end{figure}

\begin{figure}
\plotone{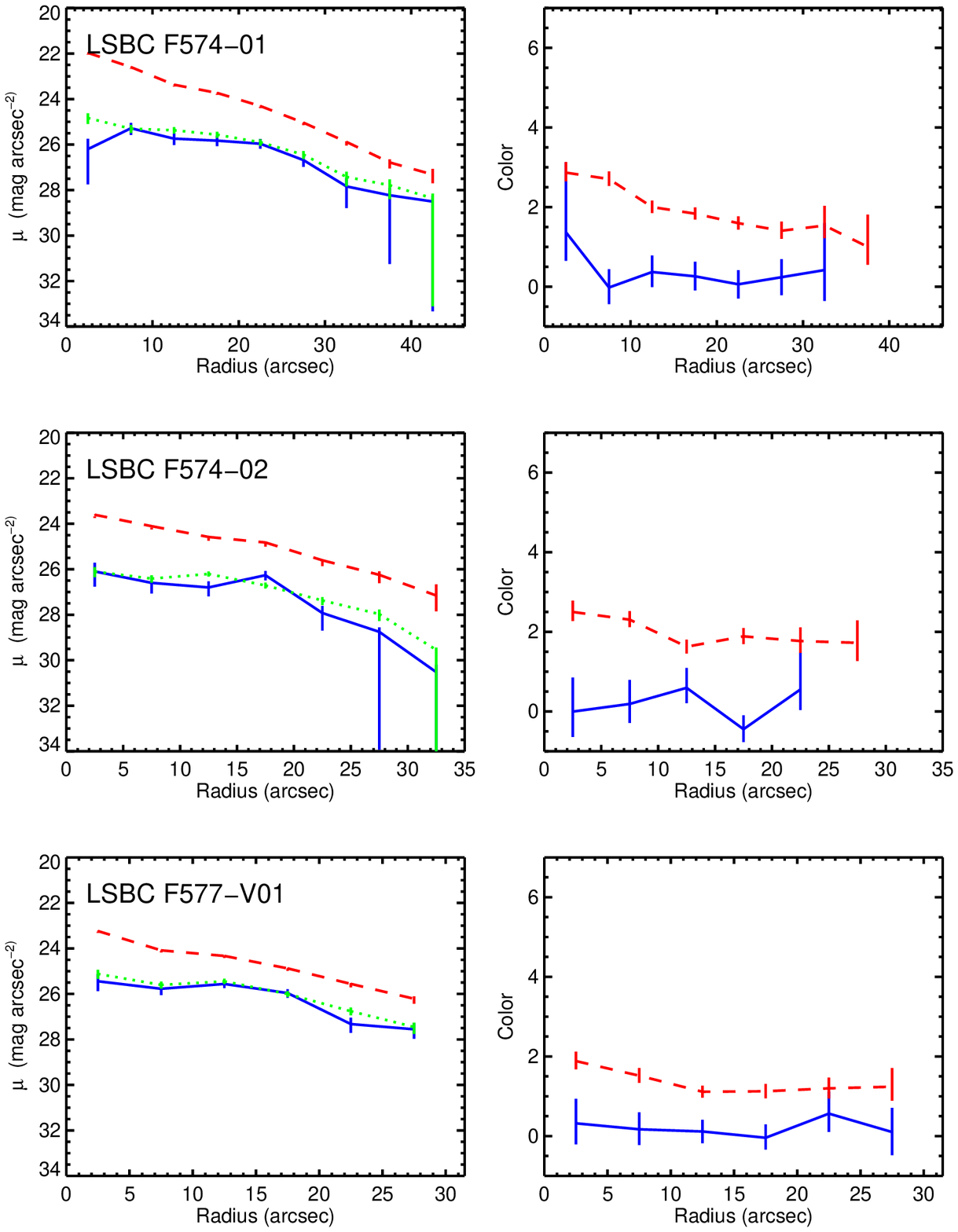}
\caption{Same as Figure \ref{deblok_radprof1}. \label{deblok_radprof4}}
\end{figure}

\begin{figure}
\plotone{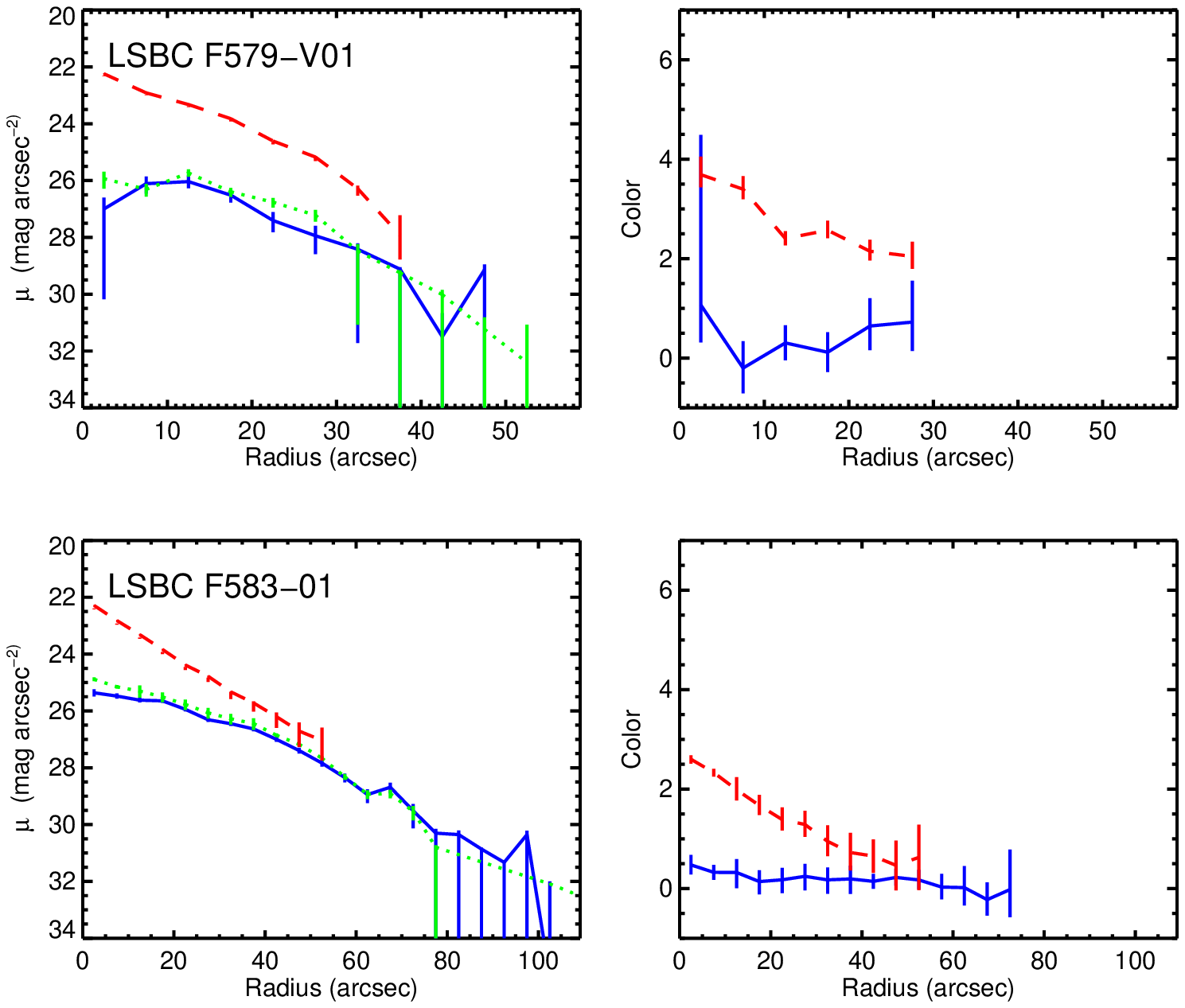}
\caption{Same as Figure \ref{deblok_radprof1}. \label{deblok_radprof5}}
\end{figure}

\begin{figure}
\plotone{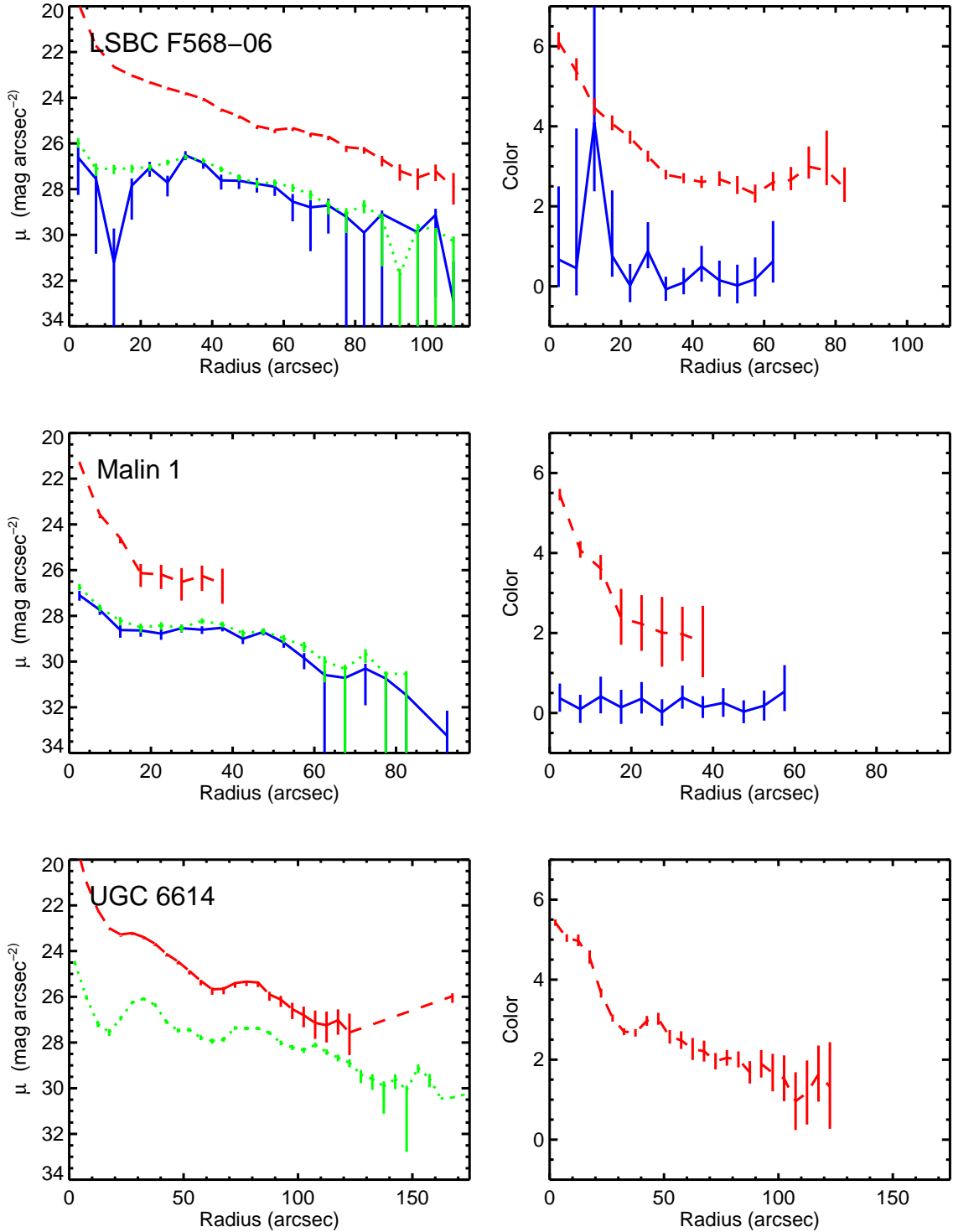}
\caption{Radial surface brightness (left) and color (right) profiles for the three giant LSB galaxies LSBC F568-06, Malin 1, and UGC 6614. The blue solid, green dotted, and red dashed curves in the left panel are the surface brightness profiles in the $FUV$, $NUV$, and $r$ bands, respectively. In the panels on the the right the $(FUV-NUV)$ color profiles are plotted as solid blue line while the dashed red curves show the $(NUV-r)$ color. 
\label{pickering_radprof}}
\end{figure}

\begin{figure}
\plotone{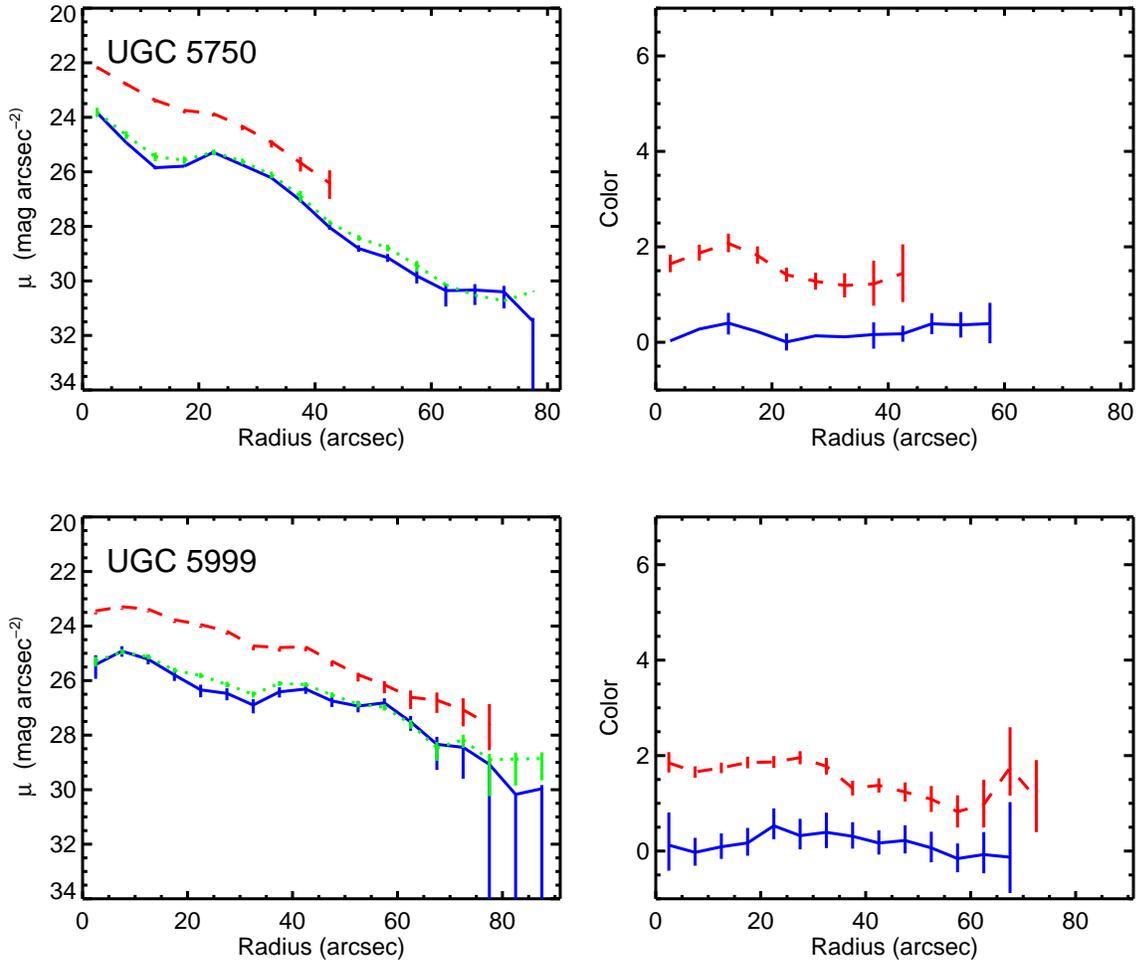}
\caption{Radial surface brightness (left) and color (right) profiles for UGC 5750 and UGC 5999. The blue, green, and red curves in the left panel are the surface brightness profiles in the $FUV$, $NUV$, and $r$ bands, respectively. In the panels on the the right the $(FUV-NUV)$ color profiles are plotted in blue while the red curves show the $(NUV-r)$ color. 
\label{vanderhulst_radprof}}
\end{figure}

In Figures \ref{deblok_radprof1}$-$\ref{vanderhulst_radprof}, we plot the radial surface brightness profiles in the panels on the left while the  color profiles are shown on the right. Most of the galaxies tend to have exponential profiles in the optical while in the UV the profiles tend to have somewhat shallower gradients. This is reflected in the $(NUV-r)$ color profiles which for many of the galaxies become bluer in the outer regions compared to the center.  On the other hand, the $(FUV-NUV)$ color does not vary significantly with radius in most of the galaxies. 

\section{Results}

In this section we use the UV measurements described in \S2 to estimate star formation rate surface densities for our sample of 19 LSB galaxies. The two main assumptions that go into transforming the UV fluxes into star formation rates are that there is little dust absorbing the UV and that the star formation history has been relatively constant. We try to constrain both of these issues using available data. Finally we combine the UV star formation rates with the \ion{H}{1} measurements from the literature to plot the LSB galaxies on the two versions of the star formation law.

\subsection{Star formation rates}

The UV light from galaxies traces directly the photospheric emission from massive stars and thus can be used to determine the SFRs of galaxies \citep[e.g.][]{kennicutt1998b}. In particular, the {\it GALEX} $FUV$ band is sensitive to stars with main sequence lifetimes shorter than $\sim10^8$ yrs \citep{martin2005}, and thus, SFRs determined from the UV are effectively averages of the SFR over this time scale. 

Based upon the data from the SINGS sample, \citet{kennicutt2007} and \citet{calzetti2007} have argued that it is best to combine a measure of the unattenated star formation rate (e.g. H$\alpha$ or UV flux) with a measure of the dust reprocessed light detected in the FIR at 24$\micron$. For our estimates of the star formation rates, we have assumed that low surface brightness galaxies have little dust, and that most of the light from recent star formation in LSB galaxies escapes unimpeded in the UV. We will test this assumption with the small amount of FIR data available for LSB galaxies in the next section.

We determined the Galactic reddening $E(B-V)$ at the position of each galaxy from the maps presented in \citet{schlegel1998} and these values are listed in Table \ref{sample}. For the \citet{cardelli1989} extinction law with $R_V=3.1$, the ratio of the extinction to the reddening is $A_{UV} / E(B-V) = 8.2$ in both of the {\it GALEX} bands \citep{wyder2007}. In addition, all of the surface brightness measurements in this paper were corrected for the $(1+z)^4$ surface brightness dimming with redshift. 

For easy comparison with previous results, we have assumed the same relation between UV luminosity and SFR as given in \citet{kennicutt1998b}:
\begin{equation}
{\rm SFR} = 1.4 \times 10^{-28} L_{\nu},
\label{sfr_uv_eqn}
\end{equation}
where SFR is the star formation rate in units of $M_{\odot}$ yr$^{-1}$ and $L_{\nu}$ is the UV luminosity in ergs s$^{-1}$ Hz$^{-1}$ measured over the wavelength range $1500-2800$ \AA,~where the UV spectrum from star-forming galaxies is expected to be relatively flat. This relation assumes a constant star formation rate and a \citet{salpeter1955} stellar initial mass function (IMF) extending from 0.1 to 100 $M_{\odot}$. The relation above can be modified to give the SFR surface density as a function of the UV surface brightness:
\begin{equation}
\log{\Sigma_{SFR}} = 7.413 - 0.4 \mu_{UV},
\label{sfr_uvsb_eqn}
\end{equation}
where $\Sigma_{SFR}$ is  the SFR surface density in $M_{\odot}$ yr$^{-1}$  kpc$^{-2}$ and $\mu_{UV}$ is the UV surface brightness in mag arcsec$^{-2}$. We used the $FUV$ surface brightness to determine $\Sigma_{SFR}$ for our sample except for UGC 6614 where we used the $NUV$ measurement and an assumed color of $(FUV - NUV) \approx 0.2$ mag due to the lack of $FUV$ data for that galaxy. When computing the $FUV$ surface brightness, we corrected the galaxies to face-on by dividing by the area of a circle with radius given by the semi-major axis of the ellipse used to extract the flux. This is equivalent to assuming that the galaxies are intrinsically circular and are optically thin in the UV. The average SFR surface densities $\Sigma_{SFR}$ calculated from equation (\ref{sfr_uvsb_eqn}), are listed in Table \ref{sflaw_table}. 

Most recent determinations of the IMF agree with the \citet{salpeter1955} form at higher masses but have relatively fewer stars below about 1 $M_{\odot}$ \citep{kroupa2002}. Thus, the conversion between UV luminosity and SFR for other IMFs would lead in general to lower SFRs than predicted by equation (\ref{sfr_uv_eqn}). If the IMF does not vary from galaxy to galaxy, then assuming a different IMF would change all of the SFRs simply by a constant factor. 

\subsection{Dust in low surface brightness galaxies}

One of the largest uncertainties in using the UV to measure SFRs is the unknown amount of light absorbed by dust internal to each galaxy. It is usually assumed that LSB galaxies have little dust although there is not very much actual data supporting this assumption. Relying upon number counts of distant galaxies observed through the giant LSB galaxy UGC 6614, \citet{holwerda2005} concluded that the $I$-band attenuation in UGC 6614 is consistent with zero, although the uncertainties as a function of radius are $\sim \pm 1$ mag. When translated to the UV, the uncertainties would be even larger and thus do not place very strong constraints on the amount of UV light absorbed by dust. \citet{matthews2001b} used a Monte Carlo radiative transfer code to model the amount of dust absorption in the edge-on LSB galaxy UGC 7321 and found that the radial color gradients were consistent with a small amount of dust reddening. According to this model, the reddening would be entirely negligible  if this galaxy were viewed closer to face-on.

There have also been several studies using the ratios of Balmer lines to determine the reddening for \ion{H}{2} regions in LSB galaxies. \citet{mcgaugh1994} calculated the reddening $E(B-V)$ from the Balmer lines, finding that some LSB galaxies have a reddening consistent with zero although their sample has values ranging up to $E(B-V) \sim 0.6$ mag. Fitting the measurements for their entire sample of LSB galaxies, \citet{burkholder2001} found an average reddening of $E(B-V) = 0.3 \pm 0.05$ mag. A similar range of reddenings  from near zero to $\sim 0.4$ (with a tail reaching to even higher values) was found by \citet{bergmann2003} and \citet{deblok1998} from the Balmer lines as well. There appear to be variations in the amount of dust even among \ion{H}{2} regions within the same galaxy. For example, the reddenings among seven individual \ion{H}{2} regions within the LSB galaxy LSBC F563-01 lie in the range $E(B-V) = 0-1$  mag \citep{deblok1998}. 

It is not clear how these measurements translate into values for the attenuation at UV wavelengths. Based upon observations of starburst galaxies, \citet{calzetti1994} found that the reddening in the ionized gas determined from the Balmer lines is about half that for the stellar continuum. Using the starburst reddening law, the ratio of $FUV$ attenuation to the reddening in the gas is $A_{FUV}/E(B-V) = 4.5$. Using the \citet{calzetti1994} ratio, the UV attenuation implied by the Balmer line data would lie in the range from zero to $\sim 2$ mag. Thus, there would appear to be a range of dust content at least within the ionized gas in LSB galaxies. On the other hand, it is not clear that the same relation between the reddening in the ionized gas and the stellar continuum for starburst galaxies would necessarily apply to the much lower density environments in LSB galaxies. 

As the UV light absorbed by dust is eventually re-emitted in the FIR, the ratio of FIR to UV flux can be used to estimate the UV attenuation nearly independently of the dust geometry or intrinsic dust properties as well as the metallicity or age of the stellar population except for galaxies with very old populations \citep{gordon2000}. Recently, there have been FIR measurements of a small sample of five LSB galaxies with the MIPS instrument aboard the {\it Spitzer Space Telescope} \citep{hinz2007, rahman2007}. Only two of these galaxies, the giant LSB galaxies Malin 1 and UGC 6614, are in common with the sample presented in this paper. However an additional two LSB galaxies with {\it Spitzer} data, namely UGC 6151 and UGC 9024, have also been observed in the UV with {\it GALEX}. We have obtained UV fluxes for these two galaxies using the same procedure as for the other LSB galaxies as described in \S2 except that we chose the axis ratio, position angle, and semi-major axis of the ellipse used to measure the total flux based upon the $NUV$ image. 

We used the formulae given in \citet{dale2002} and the observations at 24, 70, and $160 \micron$ from \citet{hinz2007} to calculate the total FIR flux from $8-1000$ \micron. Only UGC 6614 and UGC 6151 have detections in all three MIPS bands and thus we only computed total FIR fluxes for these two galaxies. Malin 1 is only detected at $24 \micron$ while UGC 9024 is detected only at 24 and $70 \micron$. In these cases, we calculated an upper limit for the total FIR luminosity using the fluxes in the detected bands plus the $3\sigma$ upper limits given by \citet{hinz2007} for the remaining bands. We calculated an $UV$ flux for each galaxy from the observed $FUV$ flux using $F_{UV} = BC_{stars} \nu F_{\nu}$, where $BC_{stars}$ is a bolometric correction from the $FUV$ flux to the total unattenuated stellar emission over $912-10000$ \AA. We assumed a value of $BC_{stars} = 1.68$ from \citet{calzetti2000}.  For UGC 6614 and UGC 9024, we calculated a $FUV$ flux using the $NUV$ measurement and assuming a color of $(FUV-NUV)=0.2$ mag. 

\begin{figure} 
\plotone{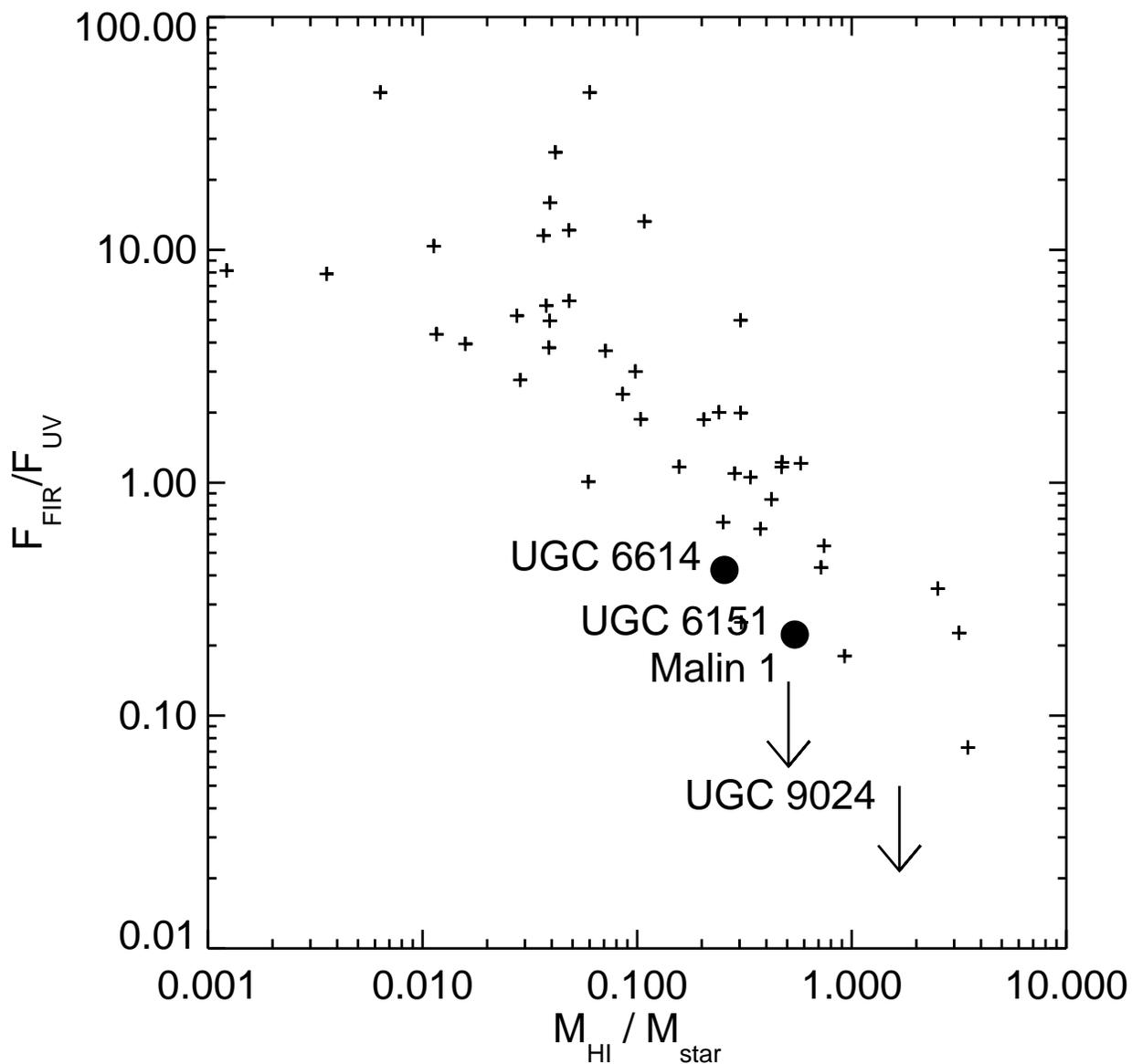}
\caption{The ratio of FIR to FUV flux as a function of the \ion{H}{1} to stellar mass ratio. The four LSB galaxies with {\it Spitzter} FIR fluxes and {\it GALEX} UV data are plotted as the large circles and upper limits. The black pluses are for the SINGS sample \citep{kennicutt2003, dale2007}. Since both UGC 6614 and UGC 6151 do not have FUV images, we have assumed a typical color of $(FUV-NUV) = 0.2$ mag.\label{irx}}
\end{figure}

The ratio of FIR to UV flux is plotted as a function of the \ion{H}{1} to stellar mass ratio $M_{HI} / M_{star}$ in Figure \ref{irx}. For a comparison sample, we plot the positions of a subset of the SINGS sample that has data in all three {\it Spitzer} MIPS bands as well as \ion{H}{1} masses and $K$-band measurements \citep{kennicutt2003, dale2007}. To determine stellar masses of the SINGS galaxies, we used the relation between $K$-band mass to light ratio and $(B-R)$ color from \citet{bell2003} converted to our assumed IMF. While the SINGS sample is not a statistical sample, it was chosen to span the parameters of normal spiral and irregular galaxies in the local universe. Thus, it provides a useful comparison sample of primarily higher surface brightness spiral and irregular galaxies. To determine the stellar masses for the LSB galaxies, we relied upon the relation between $K$-band stellar mass to light ratio and $(g-r)$ color from \citet{bell2003}. Two out of the four galaxies with FIR data have $(g-r) \approx 0.7$ while the remaining two do not have SDSS imaging. Thus we assume a color of $(g-r) =0.7$ for calculating the mass to light ratio for all four LSB galaxies. In addition, as the LSB galaxies do not have $K$-band fluxes measured, we converted the 4.5$\micron$ fluxes to $K$-band values assuming a color of $(K - [4.5]) \sim -1$ derived from the mode of the color distribution for the SINGS sample. 

The largest ratio among the LSB galaxies is $F_{FIR}/F_{UV} = 0.42 \pm 0.06$ for UGC 6614 whereas the other measurement and two upper limits are 0.2 or lower. A ratio of one would imply equal amounts of light being emitted in the UV and FIR. The fact that the ratio is significantly less than one in the LSB galaxies means that the majority of UV light from massive stars is not absorbed by dust. This stands in contrast to the majority of the SINGS galaxies which in general have FIR to UV ratios larger than one. LSB galaxies in general are relatively gas rich and this is consistent with the higher $M_{HI} / M_{star}$ ratios compared to many of the higher surface brightness galaxies in the SINGS sample. Based upon the relatively low FIR fluxes for these four galaxies, we assumed for the purposes of this paper that the UV attenuation in our sample of LSB galaxies is negligible.  However, it is important to bear in mind the relative paucity of data on the dust content of LSB galaxies.

These FIR and UV fluxes refer to the average UV attenuation within each galaxy. There are almost certainly more local variations in the FIR/UV ratio within each galaxy. For instance, comparing the UV image of UGC 6614 in Figure \ref{pickering_images} with the $24\micron$ image in \citet{hinz2007}, FIR emission is only detected in the center of the galaxy and in the inner-most star-forming ring while the UV emission extends to much larger radii. It seems likely that the FIR/UV ratio would decrease with radius, an effect not uncommon among higher surface brightness galaxies \citep{popescu2005, boissier2007}.

\subsection{Stellar populations}

Another important assumption we have made in order to use the UV fluxes to infer star formation rates for the LSB galaxies is that the UV light is coming only from massive young stars and that the star formation rate in the LSB galaxies has been relatively constant, or at least constant on the time scales sampled by stars in the UV. In this section we use the colors of the LSB galaxies to place some constraints on their stellar populations and star formation histories.

Since the UV is an indicator of the recent SFR and the optical $r$-band is more representative of the total stellar mass, the $(NUV - r)$ color provides information about the SFR divided by the stellar mass M$_*$, or the specific SFR, and hence the average age of the stars in a galaxy \citep[e.g.][]{salim2005}. While the relationship between $(NUV - r)$ color and SFR/M$_*$ is complicated by the effects of dust in high surface brightness spiral galaxies, the likely smaller amount of dust attenuation present in LSB galaxies (see \S3.2), renders interpretation of the color in terms of SFR/M$_*$ more robust. 

\begin{figure}
\plotone{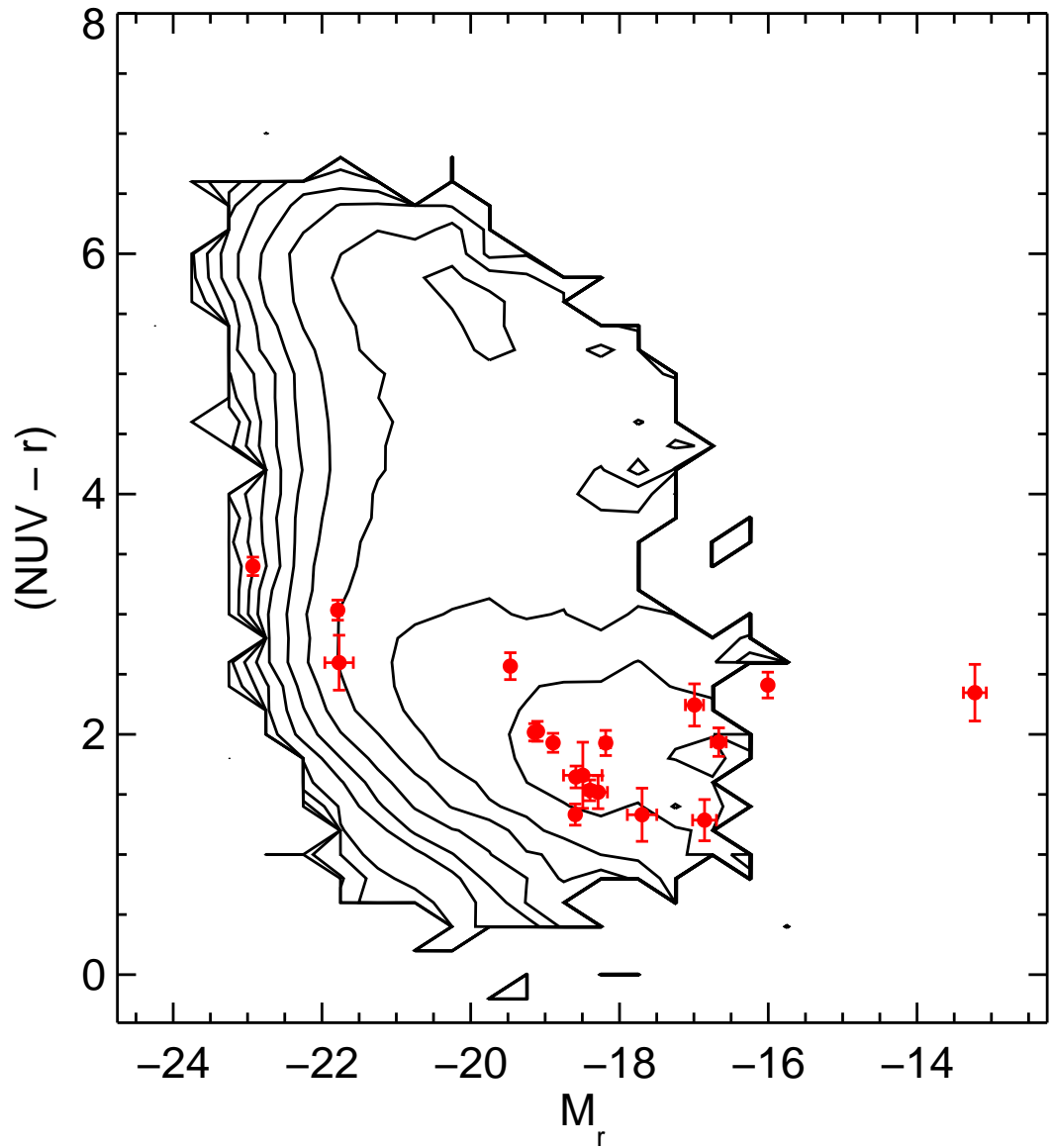}
\caption{The locations of the LSB galaxy sample in the $(NUV-r)$ vs. $M_r$ galaxy color-magnitude diagram are indicated by the solid red circles with error bars. The contours indicate the volume density of higher surface brightness galaxies from the SDSS main galaxy sample with {\it GALEX} UV measurements from \citet{wyder2007}. The contours are spaced logarithmically from $10^{-5.5}$ to $10^{-2.3}$ Mpc$^{-3}$ mag$^{-2}$. The LSB galaxies have colors similar to other higher surface brightness star-forming galaxies with similar luminosities. 
\label{cmd}}
\end{figure}

The location of the LSB galaxies in the $(NUV-r)$ vs. $M_r$ galaxy color-magnitdue diagram is shown in Figure \ref{cmd} as the red points with errors bars. For comparison we plot as contours the volume density of galaxies in the local universe from \citet{wyder2007} based upon matching the SDSS main galaxy sample with the {\it GALEX} Medium Imaging Survey catalog. While the SDSS main galaxy sample includes an explicit cut on the $r$-band half light surface brightness of $\mu_{r,1/2} < 24.5$ mag arcsec$^{-2}$ for spectroscopic target selection \citep{strauss2002}, the completeness of the SDSS sample is greater than 50\% only for $\mu_{r,1/2} < 23.4$ mag arcsec$^{-2}$ \citep{blanton2005}. Since some of the LSB galaxies in our sample have half light surface brightnesses brighter than this limit (see Table \ref{sdssphot}), there is some overlap in surface brightness between our sample and that used to generate the contours in Figure \ref{cmd}. Nevertheless, this comparison sample is dominated by higher surface brightness galaxies and can serve as a useful reference with which to compare the LSB sample. 

Galaxies tend to exhibit a bimodal distribution in color and this is reflected in the blue and red sequences visible in the color-magnitude diagram. Our sample of LSB galaxies lies along the same blue sequence as defined by the higher surface brightness galaxies. While there can be UV flux from evolved low mass stars, these galaxies have $(NUV - r) > 4$ mag, as is evident from the location of the red sequence in Figure \ref{cmd}. Thus, the colors of the LSB galaxies would imply that their UV emission is dominated by star formation and not by older stars. There does appear to be a tendency, especially for the LSB galaxies with $-20 < M_r < -17$, to be somewhat bluer than the peak of the blue sequence but the small sample size and heterogeneous selection criteria of our sample preclude us from making stronger inferences about the variation in color with surface brightness. The somewhat bluer colors would be consistent with the lower dust content in LSB galaxies. The fact that the LSB galaxies lie along the blue sequence would imply that their stellar populations are on average similar to high surface brightness galaxies of similar luminosity. Galaxies that are in the process of quenching their star formation and transitioning to the red sequence would be found at intermediate colors around $(NUV - r) \sim 4$ mag \citep{martin2007}. With the exception perhaps of Malin 1, the LSB galaxies are not found in this transition region of the color-magnitude diagram and would argue that their low surface brightness is not due to fading of a once higher surface brightness galaxy that has already begun to shut off its star formation.

Based upon N-body simulations of LSB galaxies and their ISM, \citet{gerritsen1999} found that the instantaneous SFRs in LSB galaxies fluctuate about their average value over a time scale of 20 Myr and with an amplitude of $\sim 0.1$ $M_{\odot}$ yr$^{-1}$. These flucuations are due to the finite size of the gas and star particles in their model, presumed to correspond in real galaxies to the formation of individual clusters or OB associations. If the average SFR is less than 0.1 $M_{\odot}$ yr$^{-1}$, these fluctuations would translate into a rather large spread in the colors of LSB galaxies. The integrated SFRs for our sample are listed in Table \ref{sflaw_table}. Fourteen out of the 19 galaxies in our sample have SFRs greater than 0.1 $M_{\odot}$ yr$^{-1}$ at a level where the stochastic fluctuations in the SFR would have much less of an effect on the integrated colors. Thus, our sample more closely resembles the high SFR model of \citet{gerritsen1999} which has a SFR that decreases only slightly over the 3 Gyr of time followed by their simulation. In addition, our use of the $FUV$ band to measure the SFRs would tend to average over any fluctuations in the SFR on time scales substantially less than 100 Myr.

\begin{figure}
\plotone{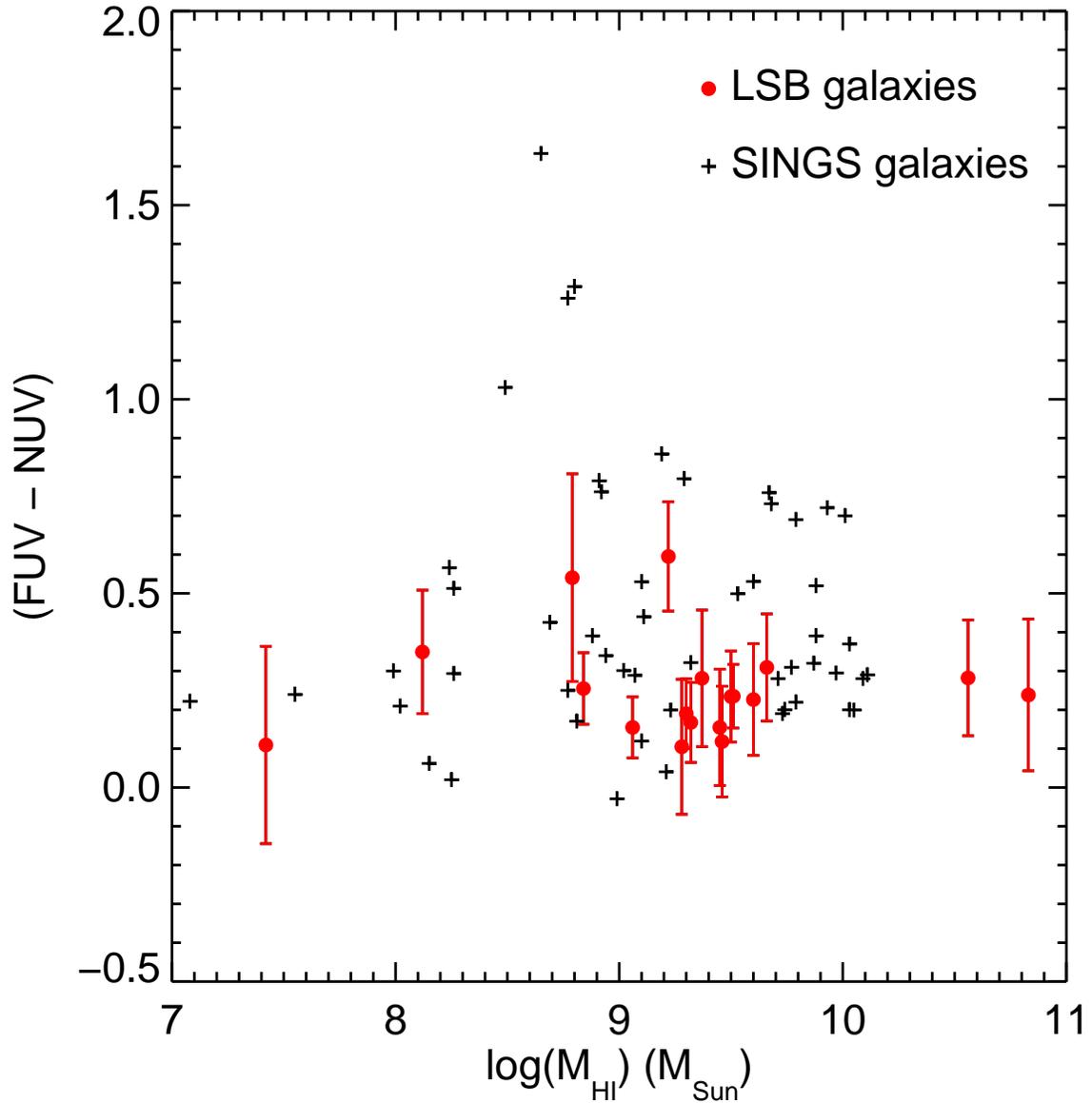}
\caption{The $(FUV - NUV)$ colors as a function of the total \ion{H}{1} mass. The LSB galaxies are plotted as the red points with error bars while the black pluses are for the SINGS galaxies \citep{kennicutt2003, dale2007}. With the exception of LSBC F563-V01 and LSBC F561-01, the LSB galaxies have fairly blue UV colors similar to most higher surface brightness star-forming galaxies.
\label{fn}}
\end{figure}

Since the $NUV$ band is sensitive to stars with a somewhat larger range of ages than the $FUV$ band \citep{martin2005}, the $(FUV-NUV)$ color can provide some constraints on the very recent star formation history ($\lesssim 1$ Gyr). In Figure \ref{fn}, we plot in red the $(FUV-NUV)$ colors of the LSB galaxies as a function of the \ion{H}{1} mass. For comparison we also plot the locations of the SINGS sample as the black pluses. With a couple of exceptions, the LSB galaxies have colors $(FUV-NUV) \sim 0.2$ mag, similar to most of the galaxies in the SINGS sample. 

Based upon a sample of 15 LSB galaxies with {\it GALEX} $FUV$ and $NUV$ data, \citet{boissier2008} found that the $(FUV-NUV)$ colors become redder above an \ion{H}{1} mass of $10^{10}$ $M_{\odot}$. On average the $(FUV-NUV)$ colors for the LSB galaxies in this paper are not as red as those presented in \citet{boissier2008}. The largest difference is for Malin 1, for which \citet{boissier2008} measured a color of $0.84\pm0.10$. In contrast we find a color of $0.24\pm 0.2$. Our measurements of the total $NUV$ magnitude agree while our $FUV$ flux is brighter than that measured by \citet{boissier2008}. This difference is due in part to the difference in the {\it GALEX} calibration between the GR1 and GR3 data releases as well as differences in the choice of aperture and the precise sky background level. As most of the galaxies in \citet{boissier2008} do not have resolved \ion{H}{1} maps, these were not included in our sample. While there does appear to be some LSB galaxies with redder UV colors, particularly at higher masses, this does not appear to be universally true. 

The $(FUV-NUV)$ colors of galaxies can be affected by several factors including the recent star formation history, the metallicity, and reddening due to dust. We have argued in \S3.2 that LSB galaxies likely have low amounts of UV attenuation from dust and therefore, dust probably does not affect the $(FUV-NUV)$ colors of our sample. There is some disagreement among stellar population models about the intrinsic $(FUV-NUV)$ colors of galaxies. For models with a constant star formation rate, no dust, solar metallicity, and a \citet{kroupa1993} stellar IMF reaching to 100 $M_{\odot}$, \citet{boissier2008} predict a color of $(FUV-NUV) \approx 0.2$ mag after about 1 Gyr. Models with lower metallicities yield slightly bluer colors. On the other hand, the models of \citet{bruzual2003} for a similar set of parameters predict a color of $(FUV-NUV) \approx 0.0$ mag for ages greater than 1 Gyr. With only a couple of exceptions, the $(FUV-NUV)$ colors of all of our LSB galaxies are consistent to within the errors with the Boissier et al. models and somewhat redder than that predicted by Bruzual \& Charlot. Given the errors on the $(FUV-NUV)$ color and the disagreement among models, we do not find any strong evidence from the colors of the LSB galaxies for either variable star formation histories or a non-standard IMF. If this had been the case, then we would be underestimating the SFRs in the LSB galaxies using the standard conversion factor between UV luminosity and SFR in equation (\ref{sfr_uv_eqn}).

\subsection{Gas surface densities}

We have used the \ion{H}{1} radial surface density profiles from \citet{vanderhulst1993}, \citet{deblok1996}, and \citet{pickering1997} to measure average gas surface densities for our sample of LSB galaxies within the same aperture used to measure the total UV flux. In order to be consistent with measurements from \citet{kennicutt1998a}, we did not correct the gas densities for helium or other heavy elements.

The total gas surface density should include both the atomic and molecular gas. Only a few LSB galaxies have molecular gas detected from radio observations of the CO lines while most remain undetected \citep{deblok1998b, oneil2000, matthews2001a, oneil2003a, oneil2004, matthews2005, das2006, schombert1990, braine2000}. The few detections and many upper limits correspond to very low molecular fractions in the range $1-10\%$ for most LSB galaxies, assuming a Galactic CO to H$_2$ conversion factor \citep{oneil2003a}. Among the galaxies with molecular gas detected, CO maps of the giant LSB galaxies LSBC F568-06 and UGC 6614 show molecular gas clearly offset from the nucleus and only detected at certain locations, indicating that what little molecular gas they do have is irregularly distributed \citep{das2006}.  

One critical assumption that went into determining these low molecular fractions is that the standard Galactic conversion factor between CO luminosity and H$_2$ mass applies to LSB galaxies. Since LSB galaxies have on average oxygen abundances below the Solar value \citep{burkholder2001, mcgaugh1994}, the ratio of CO to H$_2$ would be expected to be lower simply due to the overall lower metallicity. On the other hand, observations of individual molecular clouds in nearby low metallicity dwarf galaxies are consistent with the standard Galactic CO-to-H$_2$ conversion factor \citep{leroy2006, bolatto2008}. In addition, the low dust content in LSB galaxies would be expected to lower the CO/H$_2$ ratio because the dust can act as a catalyst for the formation of CO as well as shielding the molecules from potentially damaging UV radiation \citep{mihos1999}. Despite these uncertainties, we assumed for the purposes of this paper that the gas mass in LSB galaxies is dominated by the atomic gas.

\subsection{The star formation law}

\begin{figure}
\plotone{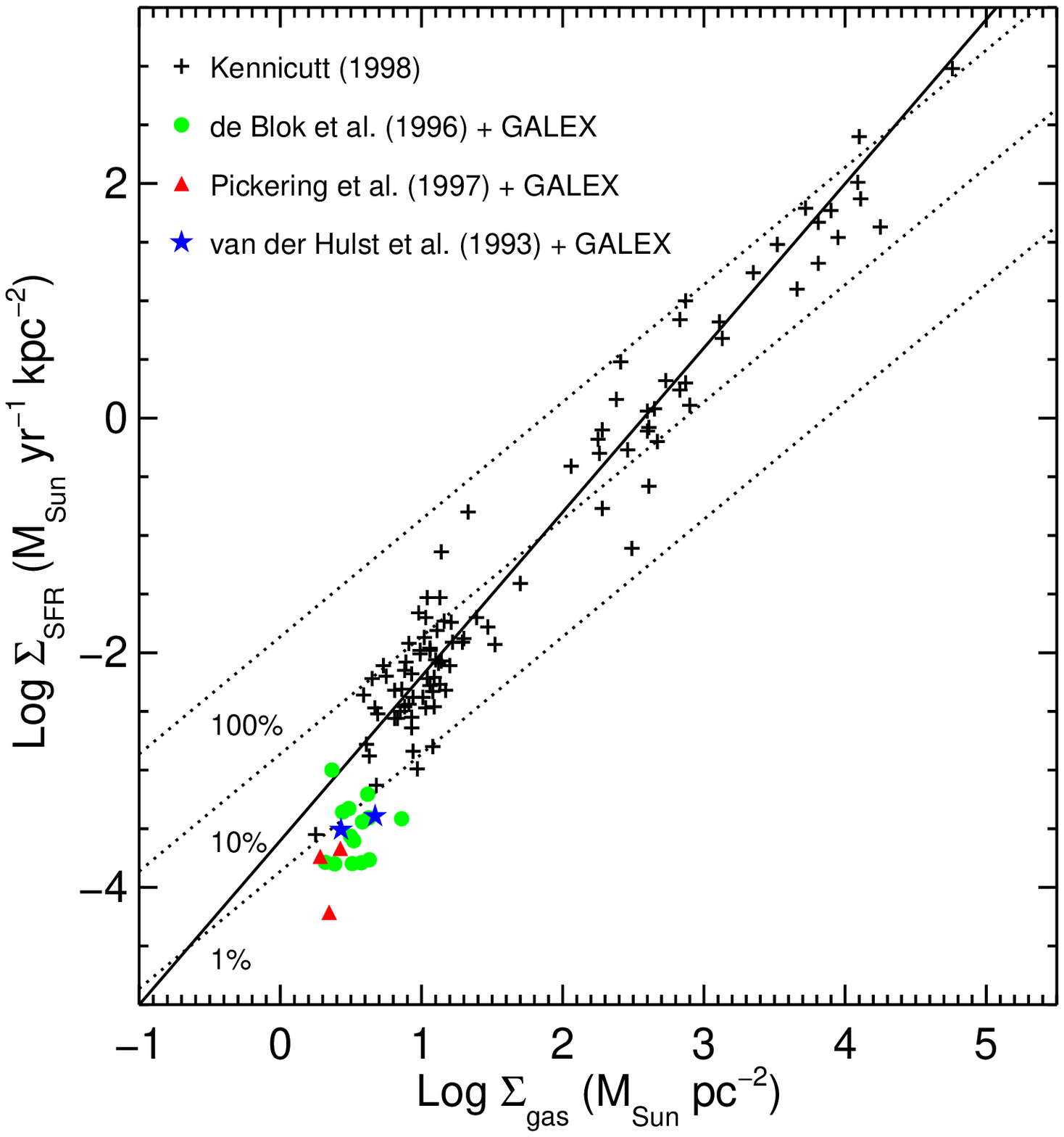}
\caption{The SFR surface density as a function of the total hydrogen gas surface density. The colored symbols indicate the sample of 19 LSB galaxies from this paper with SFRs measured from the UV with no correction for dust attenuation. The gas surface densities are derived from the \ion{H}{1} data from \citet{deblok1996} (green circles), \citet{pickering1997} (red triangles), and \citet{vanderhulst1993} (blue stars) and assume the molecular fraction is negligible. The black pluses indicate the sample of higher surface brightness galaxies from \citet{kennicutt1998a} while the solid line is the power law fit to these points with exponent 1.4. The dotted lines indicate lines of constant star formation efficiency assuming a star formation time scale of $10^8$ yrs. The LSB galaxies tend to lie below the extrapolation of the power law fit to the higher surface brightness sample.
\label{sflaw1}}
\end{figure}

We plot the star formation rate surface density as a function of the gas surface density in Figure \ref{sflaw1}. The green circles, red triangles and blue stars are the galaxies with \ion{H}{1} data from \citet{deblok1996}, \citet{pickering1997}, and \citet{vanderhulst1993}, respectively. For comparison we also plot the sample of spiral and starburst galaxies from \citet{kennicutt1998a} as the black pluses. The solid line is the power law fit to the high surface brightness sample of the form of equation (\ref{sflaw1_eqn}). \citet{kennicutt1998a} found best fit values of $A = (2.5 \pm 0.7) \times 10^{-4}$ and $N = 1.4 \pm 0.15$ for $\Sigma_{SFR}$ in $M_{\odot}$ yr$^{-1}$ kpc$^{-2}$ and $\Sigma_{gas}$ in $M_{\odot}$ pc$^{-2}$. The LSB galaxies lie below the extrapolation of this fit to the higher surface density galaxies. The median offset between the LSB galaxies and the extrapolation of the power law is about 0.7 dex, or a factor of five in SFR surface density.

For reference the dotted lines in Figure \ref{sflaw1} indicate lines of constant star formation efficiency of 1, 10, and 100\% as labeled in the figure assuming a time scale for star formation of $10^8$ yrs. The choice of $10^8$ yrs is somewhat arbitrary although it does correspond to typical orbital time scales in galaxies. These curves would imply that the average star formation efficiency in LSB galaxies is only at most a few percent, significantly lower than in higher surface density galaxies. 

Note that we assumed that the LSB galaxies have negligible amounts of molecular gas when computing $\Sigma_{gas}$. If in fact there were more molecular gas than would be indicated by the available CO measurements, then the LSB galaxies would shift to the right, or even further from the fit to the higher surface brightness galaxies. As described above, we have not included any correction for internal dust in the LSB galaxies. Any correction for dust would tend to push the points back closer to the fit to the higher surface brightness galaxies. However, the average correction necessary would corresponds to an internal extinction of about 1.8 mag, significantly larger than the UV attenuation of $ \lesssim 0.4$ mag implied by the FIR/UV ratios in Figure \ref{irx}. 

Another critical assumption used to calculate the SFRs in Figure \ref{sflaw1} is that the stellar IMF is the same at low and high surface densities. Recently it has been suggested that the IMF at low density should have fewer high mass stars than at higher densities \citep{krumholz2008}. If this were to be the case, then the conversion between UV luminosity and SFR in equation (\ref{sfr_uv_eqn}) would tend to underestimate the true SFR. Indeed, the extreme outer disks of some galaxies have an apparent edge in the H$\alpha$ images but no apparent edge in the UV. Since only the very most massive stars are capable of ionizing hydrogen and producing significant H$\alpha$-emitting regions, a relative lack of massive stars would help explain the low H$\alpha$/UV ratio in the outer regions of some disk galaxies \citep{thilker2005, boissier2007}. The surface densities probed in the extended disks of high surface brightness galaxies are in general in the regime below $\sim 10$ $M_{\odot}$ pc$^{-2}$, similar to the LSB galaxies in our sample. Unfortunately, there are very few integrated H$\alpha$ fluxes available for LSB galaxies so that we cannot check whether the UV/H$\alpha$ ratio is similar to that in the extended low density outer regions of high surface brightness galaxies. Nevertheless, variations in the high mass IMF would affect SFRs determined in the UV less than those from H$\alpha$ because of the wider mass range of stars contributing to the UV luminosity.

There is evidence that the outer low density regions of higher surface brightness galaxies also tend to show a steeper slope in the Schmidt-Kennicutt Law. In particular, \citet{kennicutt1989, kennicutt1998a} used radial H$\alpha$ and gas density profiles to show that galaxies tend to have well-defined edges to their star-forming disks below a gas density threshold that varies from galaxy to galaxy in the range $3-10$ $M_{\odot}$ pc$^{-2}$. Much below the threshold no H$\alpha$ emission is detected while near the threshold the correlation of SFR surface density with gas density is steeper than the 1.4 power law at higher densities. \citet{kennicutt1989} showed that the gas density thresholds could be explained as the radius inside of which the gas is unstable to gravitational instabilities as predicted by the Toomre criterion \citep{toomre1964}. LSB galaxies have gas densities below the critical density defined by the Toomre criterion throughout much of their disks \citep{vanderhulst1993}, similar to the outer regions of high surface brightness galaxies. 

\begin{figure}
\plotone{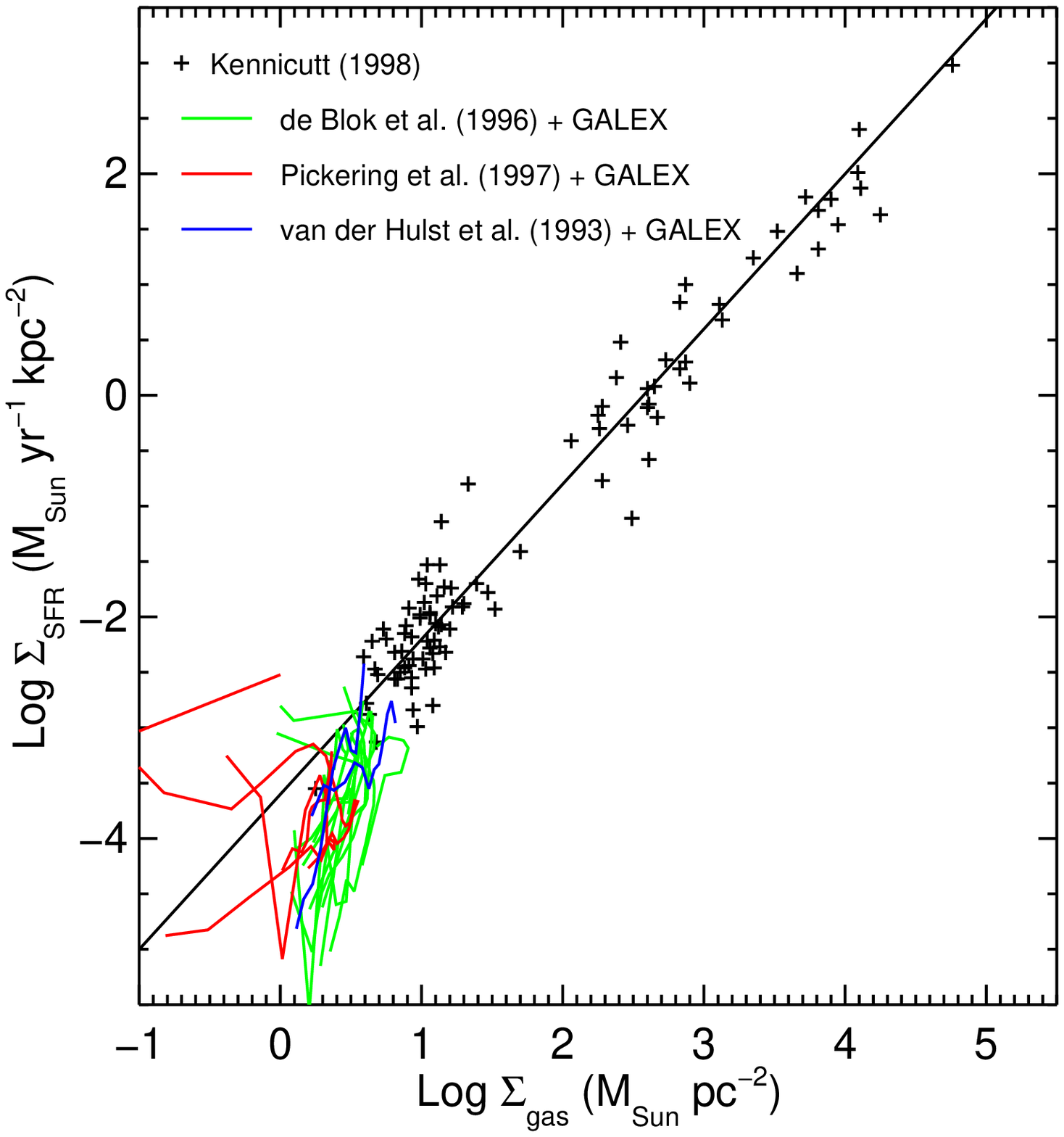}
\caption{Similar to Figure \ref{sflaw1} except that the azimuthally averaged SFR surface density is plotted as a function of the gas surface density for the LSB galaxy sample. There is a single solid line connecting the data for each individual galaxy. The lines are color-coded by the source of the \ion{H}{1} data: green for \citet{deblok1996}, red for \citet{pickering1997}, and blue for \citet{vanderhulst1993}.
The resolved profiles plotted in this figure generally lie in the same part of the diagram as the integrated measurements in Figure \ref{sflaw1}. The two exceptions are the giant LSB galaxies LSBC F568-06 and UGC 6614 (plotted in red) which tend to deviate from the main trend in their central regions.
\label{sflaw1_local}}
\end{figure}

We have used the \ion{H}{1} and UV radial profiles to investigate the form of the star formation law locally within the LSB galaxies. The azimuthally averaged gas and SFR surface densities for the LSB galaxies are plotted in Figure \ref{sflaw1_local}. A solid line connects all of the data for a particular galaxy. The colors correspond to the source of the \ion{H}{1} data as indicated in the figure. With a couple of exceptions, the azimuthally averaged data follow the same trends as seen in the galaxy-wide averaged data. The radial profiles indicate that the star formation law is steeper at low density than the canonical power law fit to the higher density points. The two red curves that deviate the most from the average trend are the giant LSB galaxies LSBC F568-06 and UGC 6614. Both of these galaxies have central minima in their \ion{H}{1} profiles even though their UV surface brightness profiles continue to increase towards their centers. Both of these galaxies have an active nucleus exhibiting broad emission lines \citep{bothun1990, schombert1998} and thus at least some of the UV emission in their centers is probably not due to star formation. The lack of \ion{H}{1} in the center could be an indication that there is a significant amount of molecular gas in the inner regions. Although the CO emission detected in these two galaxies by \citet{das2006} is in the central regions of both systems, the relative amount of gas detected would only contribute $\sim 0.2-0.3$ $M_{\odot}$ pc$^{-2}$ when azimuthally averaged over the radii covered by the CO observations and hence would not significantly affect the total gas surface density. 

Despite the downturn observed in the radial profiles in Figure \ref{sflaw1_local}, the UV and \ion{H}{1} data are still of low enough resolution such that we are averaging over fairly large regions within each galaxy. It could be that there is a constant gas density threshold and that the filling factor of gas clouds that lie above the threshold is simply lower in LSB galaxies. This would lead to an effectively lower star formation efficiency when averaging over the entire galaxy. Higher resolution UV and \ion{H}{1} data would be necessary to investigate this possibility. 

Although the correlation of $\Sigma_{SFR}$ with $\Sigma_{gas}$ has been one of the most widely used star formation prescriptions, there are alternative formulations. In particular, \citet{kennicutt1998a} showed that a relation of the form of equation (\ref{sflaw2_eqn}) fits the data for spiral and starburst galaxies as well as the power law in equation (\ref{sflaw1_eqn}). We have used the published \ion{H}{1} rotation curves for our galaxies to determine the rotational velocity $V_{rot}$ at a radius 0.7 times the radius used to determine both $\Sigma_{SFR}$ and $\Sigma_{gas}$. This choice of radius was made to make our measurements as consistent as possible with those for the higher surface brightness galaxies where the dynamical time was determined at a radius of $0.7 R_{25}$ (R. Kennicutt, private communication). For many of the galaxies in our sample, this radius occurs in the flat part of the rotation curve, and thus $V_{rot}$ is fairly well defined. Following \citet{kennicutt1998a}, we defined the dynamical time as $\tau_{dyn}  = 2 \pi R / V_{rot}$. The values for $V_{rot}$ and $\tau_{dyn}$ are listed in Table \ref{sflaw_table}. 

\begin{figure}
\plotone{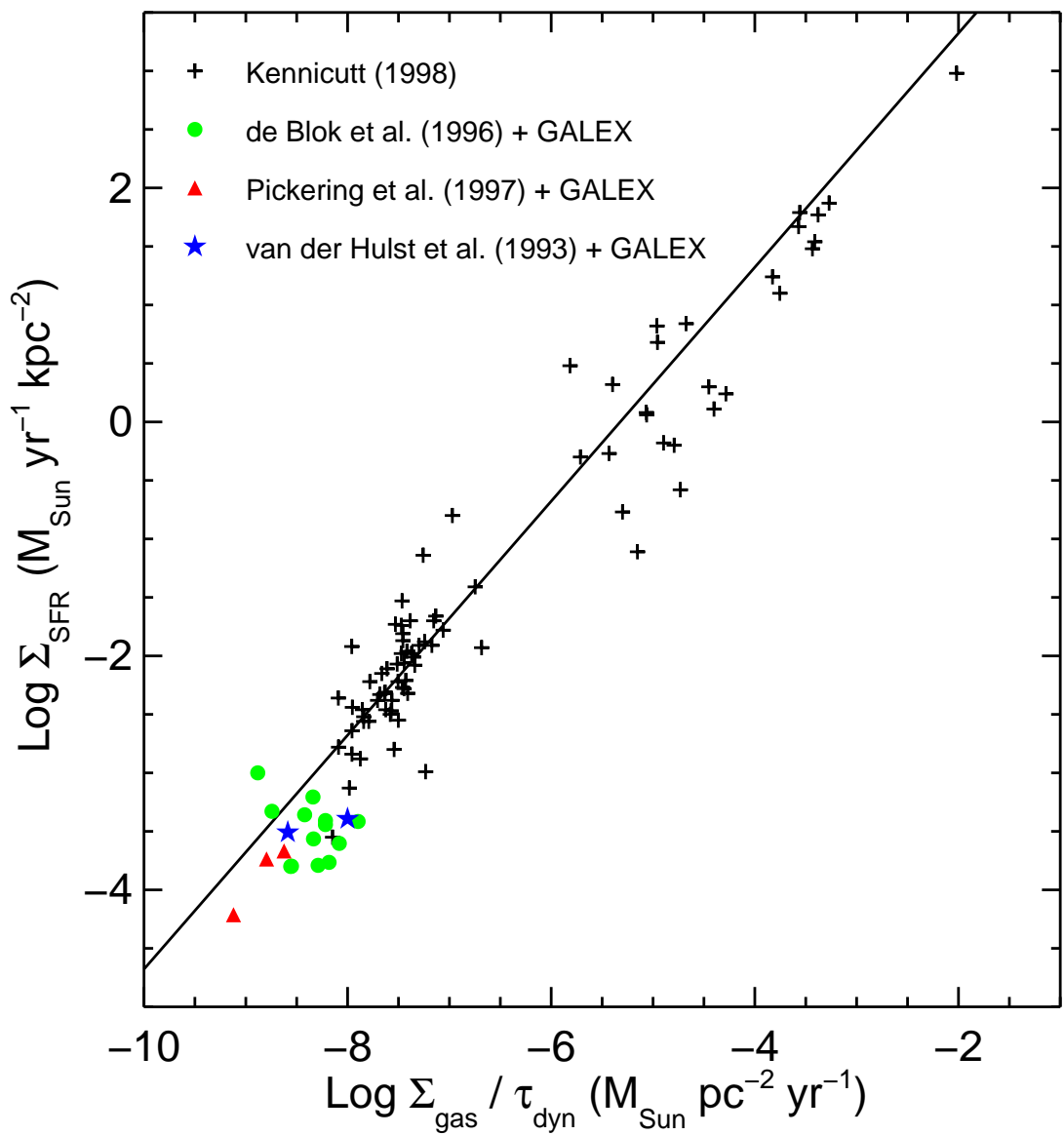}
\caption{Similar to Figure \ref{sflaw1}, except the SFR surface densities are plotted as a function of the gas surface density divided by the dynamical time. The plot symbols are the same in as in Figure \ref{sflaw1}. The solid line is the linear fit to the high surface brightness sample from \citet{kennicutt1998a} which corresponds to transforming about 11\% of the gas per orbital time scale into stars. 
\label{sflaw2}}
\end{figure}

This alternative star formation law is plotted in Figure \ref{sflaw2}. Similar to Figure \ref{sflaw1}, the LSB galaxies are plotted as the colored points with the color indicating the source of the \ion{H}{1} data as noted in the legend. The black pluses are the high surface brightness spiral and starburst sample from \citet{kennicutt1998a} while the solid black line is the fit to these points of the form of equation (\ref{sflaw2_eqn}) with $B=0.11$. Similar to the first version of the star formation law, the LSB galaxies lie below the extrapolation of the fit to the higher surface brightness sample. The median offset is 0.5 dex below although there is more scatter in the LSB SFR surface densities vs. $\Sigma_{gas} / \tau_{dyn}$ as compared to that seen in Figure \ref{sflaw1} vs. $\Sigma_{gas}$ alone.

As noted by \citet{kennicutt1998a}, one interpretation of the fit to the high surface brightness galaxies shown in Figure \ref{sflaw2} is that a constant fraction of the available gas is transformed into stars per orbit. The zeropoint of the relation plotted in Figure \ref{sflaw2} corresponds to approximately 11\% of the gas transformed into stars per orbit. This could in principle arise if star formation were triggered by passages through spiral arms in which case the star formation rate would naturally correlate with the amount of gas available and vary inversely with the orbital time. In this interpretation, while the star formation efficiency is approximately a constant among the high surface brightness galaxies, the efficiency would be somewhat lower in LSB systems.

One possible explanation for the decreased star formation efficiency at low density is the low fraction of molecular gas observed in environments where $\Sigma_{gas}  <  10 M_{\odot}$ pc$^{2}$. \citet{blitz2006} have argued that the molecular fraction in disk galaxies is correlated with the hydrostatic pressure in the ISM. In this model the pressure is a function of the stellar surface density, the gas surface density, and the gas velocity dispersion. Thus, the LSB galaxies would be expected to have lower ISM pressure and thus low molecular fractions. If it is really just the molecular gas which is relevant for star formation in galaxies, then the apparently low star formation efficiency in the total gas in LSB galaxies is simply a reflection of their low molecular content. 

\citet{krumholz2005} have developed a model in which star formation in galaxies occurs in the densest sub-regions of molecular clouds that are supersonically turbulent with a log-normal density distribution. They coupled this model with the correlation between the molecular fraction and the ISM pressure from \citet{blitz2006} to derive an analytical prediction for both forms of the star formation law. For the first form of the law as a function of $\Sigma_{gas}$, the \citet{krumholz2005} model prediction is given by
\begin{equation}
\Sigma_{SFR} = 3.9 \times 10^{-4} \phi_{\bar{P}}^{0.34} Q^{-1.32} f_{GMC} \Sigma_{gas}^{1.33}~M_{\odot}~{\rm yr^{-1}~kpc^{-2}},
\label{sflaw1_model_eqn}
\end{equation}
where $\Sigma_{gas}$ is the total (atomic plus molecular) gas surface density in units of $M_{\odot}$ pc$^{-2}$, $f_{GMC}$ is the fraction of gas in giant molecular clouds, $\phi_{\bar{P}}$ is the ratio of the pressure inside a molecular cloud to the pressure at its surface, and $Q$ is the stability parameter as defined by \citet{toomre1964}. We adopt $Q=1.5$, a value typical of spiral galaxies \citep{martin2001}. Krumholz \& McKee argued that $\phi_{\bar{P}} \approx 10-8 f_{GMC}$ and we adopt $\phi_{\bar{P}} = 9 f_{GMC}$. Finally, $f_{GMC}$ is given by
\begin{equation}
f_{GMC} = \left( 1 + 250 \Sigma_{gas}^{-2} \right)^{-1}.
\label{fgmc1_eqn}
\end{equation}
Similarly, the Krumholz \& McKee prediction for the second form of the star formation law in terms of the dynamical time is given by
\begin{equation}
\Sigma_{SFR} = 5.5 \times 10^4 \phi_{\bar{P}}^{0.34} Q^{-1.32} f_{GMC} \left(\frac{\Sigma_{gas}}{\tau_{dyn}}\right)^{0.89}~M_{\odot}~{\rm yr^{-1}~kpc^{-2}},
\label{sflaw2_model_eqn}
\end{equation}
where $\tau_{dyn}$ is in units of yrs and $f_{GMC}$ is given as a function of $\Sigma_{gas}/\tau_{dyn}$ by
\begin{equation}
f_{GMC} = \left[ 1 + 2.0 \times 10^{-10} \left(\frac{\Sigma_{gas}}{\tau_{dyn}}\right)^{-1.34} \right]^{-1}.
\label{fgmc2_eqn}
\end{equation}
Note that equation (\ref{fgmc2_eqn}) was derived from equation (\ref{fgmc1_eqn}) and an empirical relation between $\Sigma_{gas}$ and $\tau_{dyn}$.

\begin{figure}
\plottwo{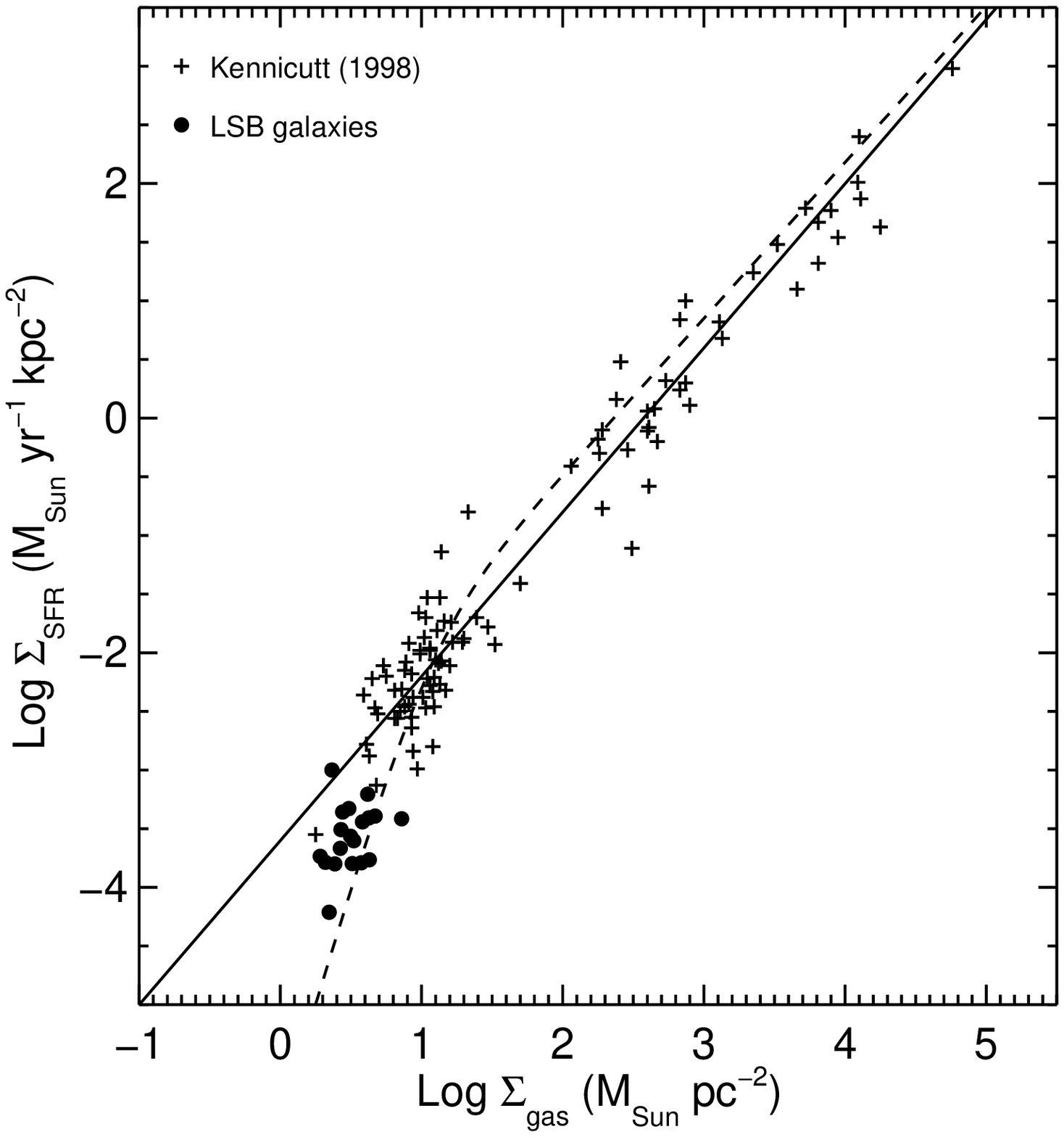}{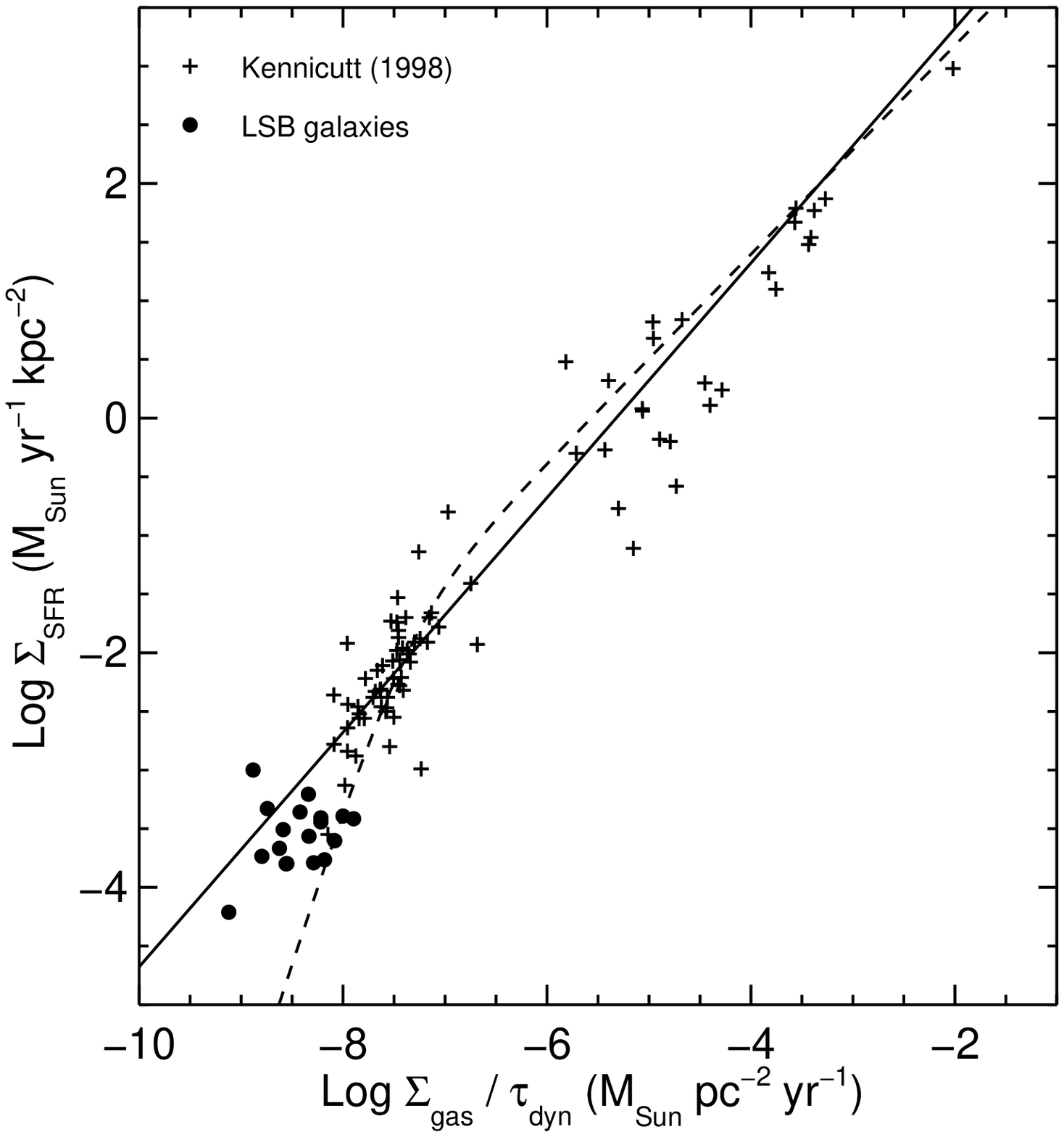}
\caption{The two versions of the star formation law as a function of $\Sigma_{gas}$ (left) and $\Sigma_{gas}/\tau_{dyn}$ (right). The circles indicate the LSB galaxies from this paper while the pluses are the spiral and starburst galaxy sample from \cite{kennicutt1998a}. The solid lines are the fits to the higher surface brightness sample and dashed lines indicate the model predictions from \citet{krumholz2005}.  
\label{sflaw_model}}
\end{figure}

In Figure \ref{sflaw_model} we plot the prediction from equations (\ref{sflaw1_model_eqn}) and (\ref{fgmc1_eqn}) as a function of $\Sigma_{gas}$ as the dashed line in the left panel. The corresponding prediction from equations (\ref{sflaw2_model_eqn}) and (\ref{fgmc2_eqn}) as a function of $\Sigma_{gas} / \tau_{dyn}$ is shown in the right panel. For both forms of the star formation law, we additionally shifted the curves to the left (i.e. lower gas densities) by a factor of 1.37 to account for the fact that the observations do not include a correction for Helium or other heavy elements.  In both forms of the star formation law the down-turn in $\Sigma_{SFR}$ at low density is due to the declining molecular fraction as a function of density, similar to what is observed in the LSB galaxies. In both panels the LSB data appear to be a bit above the theoretical prediction. However given the uncertainties in the molecular content of the LSB galaxies as well as the uncertainties inherent in deriving SFRs, the models are in reasonable agreement with the observations.

To further test the idea that is is the amount of molecular gas which determines the star formation rates in galaxies rather than the total gas density, we would require measurements of the molecular gas content in LSB galaxies. Unfortunately, only one of the galaxies in our sample has a CO detection while an additional seven galaxies have only upper limits to their CO flux \citep{das2006, schombert1990, braine2000, deblok1998b}. The CO data for these galaxies typically come from single dish measurements that sample the central regions with diameters in the range of $22-55\arcsec$, or significantly smaller than the UV and \ion{H}{1} sizes listed in Table \ref{photaper}. Therefore, we have remeasured the $FUV$ surface brightness using the same aperture as used for the CO data. The corresponding values for $\Sigma_{SFR}$ determined from equation (\ref{sfr_uvsb_eqn}) are listed in Table \ref{sflaw_mol_table}. We have converted the detection and upper limits in the CO brightness temperature into molecular hydrogen surface densities using the Galactic conversion factor 4.5 $M_{\odot}$ pc$^{-2}$ (K km s$^{-1}$) \citep{bloemen1986}, and these values for $\Sigma_{H2}$ are listed in Table \ref{sflaw_mol_table} as well.  The sole detection of CO flux is for UGC 6614, which when averaged over the $100\arcsec$ diameter region sampled by the CO data, corresponds to a surface density of $\Sigma_{H2} = 0.2$ $M_{\odot}$ pc$^{-2}$ \citep{das2006}.\footnote{Although the galaxy LSBC F568-06 was also detected in CO by \citet{das2006}, the data do not cover the full range of velocities in the disk of this object and thus cannot be used to measure a value for $\Sigma_{H2}$.}

\begin{figure}
\plotone{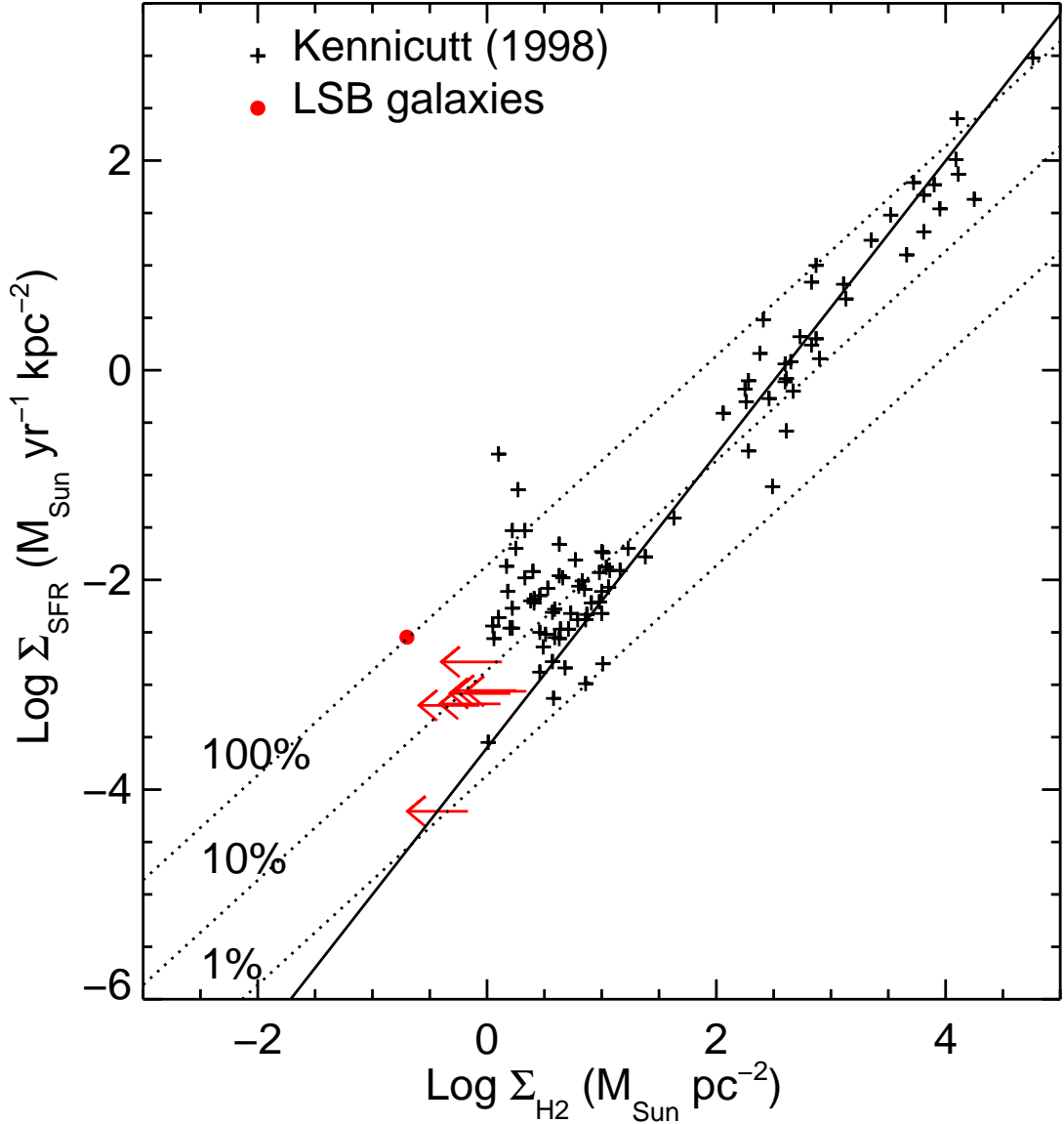}
\caption{The molecular star formation law. The star formation rate surface density is plotted as a function of the molecular gas surface density. We plot in red the locations of the subset of LSB galaxies with CO observations. The solid circle indicates the only galaxy (UGC 6614) with a CO detection while the arrows indicate the remaining galaxies with only upper limits for their CO flux. The CO data were converted into molecular gas masses using a CO/H$_2$ conversion factor of 4.5 $M_{\odot}$ pc$^{-2}$ (K km s$^{-1}$)$^{-1}$. The solid line is the 1.4 power law fit to the star formation law as a function  of the total gas surface density from \citet{kennicutt1998a}. The dotted lines show locations of constant star formation efficiency per $10^8$ yrs for efficiencies of 1\%, 10\%, and 100\%, as labeled. 
\label{sflaw_mol}}
\end{figure}

We plot $\Sigma_{SFR}$ as a function of the molecular gas surface density $\Sigma_{H2}$ in Figure \ref{sflaw_mol}.  The sole galaxy with a molecular gas detection, UGC 6614, is plotted as a red circle while red arrows indicate upper limits for the remaining seven LSB galaxies with CO observations but no detected molecular gas.  As in the previous figures, we also plot the spiral and starburst sample from \citet{kennicutt1998a}. For reference we plot the same 1.4 power law fit to $\Sigma_{SFR}$ vs. $\Sigma_{gas}$ as shown in Figure \ref{sflaw1}. While this power law fits the starburst galaxies with the highest surface densities in which the gas is assumed to be all molecular, the spiral and LSB galaxies tend to scatter to the left of this line. As noted by Kennicutt, the scatter in $\Sigma_{SFR}$ vs. $\Sigma_{H2}$ is larger than when plotting $\Sigma_{SFR}$ vs. the total atomic plus molecular gas density. It also is apparent from this plot that a  single power law does not appear to provide a very good fit to the combined sample. For reference we plot in Figure \ref{sflaw_mol} lines of constant star formation efficiency per $10^8$ yrs. The implied star formation efficiency for UGC 6614 is nearly 100\% per $10^8$ yrs while the lower limits for the CO non-detections lie in the range of $1-10$\%. Part of the increased scatter in this diagram may be due to variations in the CO to H$_2$ conversion factor. Clearly more data on the variation in this conversion factor among galaxies is needed before concluding that the scatter is induced by a real variation in star formation efficiency. 
 
Finally we note that most theoretical explanations for the star formation law indicate that the fundamental relationship is between the SFR and gas {\it volume} densities whereas we can only measure {\it surface} densities. The down-turn in both forms of the star formation law could be due to low surface brightness galaxies being on average thicker than higher surface brightness galaxies. Indeed, in their hydrodynamical simulations of star formation in galaxies \citet{robertson2008} found that flaring of the disk in the outer regions of their simulated galaxies leads to deviations from the canonical power law star formation relation with exponent of 1.4. For a disk in which the gas volume density falls off exponentially with height above the plane with scale height $h$, the gas surface density $\Sigma_{gas}$ and the mid-plane volume density $\rho_{gas,0}$ are related by $\rho_{0,gas} = \Sigma_{gas}/(2h)$. This would imply that the LSB galaxies have vertical gas scale heights a factor of five above that for higher surface brightness galaxies if the down-turn were due entirely to this effect. Optical images of the edge-on LBS galaxy UGC 7321 show that the vertical scale height of the disk is only 140 pc, a value smaller than in most higher surface brightness galaxies \citep{matthews2000}. If this is representative of LSB galaxies generally, then an increased vertical scale height would not be able to account for the low gas surface densities in LSB galaxies. Based upon N-body-SPH simulations \citet{kaufmann2007} have argued that dwarf galaxies with rotational velocities of $\sim 40$ km s$^{-1}$ are formed with significantly thicker disks than higher mass galaxies. This increased puffiness would lead to an effectively lower star formation efficiency in dwarf galaxies. While a few of the LSB galaxies in this paper have rotational velocities in this regime, the galaxies in our sample range in rotational velocity up to nearly 300 km s$^{-1}$. While it seems unlikely that all LSB galaxies have disks a factor of five more extended than high surface brightness galaxies, variations in the disk thickness could account for some of the scatter observed at low density in both forms of the star formation law. 
 
\section{Conclusions}

In this paper we have collected a sample of 19 LSB galaxies with resolved \ion{H}{1} maps in the literature, UV data from {\it GALEX}, and optical images from the SDSS in order to extend the star formation law to even lower densities than previously observed. These LSB galaxies span a wide range of luminosities ($-13 > M_r > -23$) from dwarfs to giant LSB galaxies. The UV emission in most of the galaxies is detected out to similar radii as the \ion{H}{1}, and in general extends to larger radii than observed in the optical images. 

All of the LSB galaxies have $(NUV-r)$ colors similar to other star-forming galaxies of similar total luminosity.  The fact that they lie along the same blue sequence as for higher surface brightness galaxies means that the UV emission in the LSB galaxies is dominated by light from young stars rather than being due to evolved low mass stars. With the possible exception of the giant LSB galaxy Malin 1, none of the galaxies in our sample lie in the transition region between the blue and red sequence and thus truncation of their star formation does not explain their low surface brightnesses. Since the LSB galaxies lie along the blue sequence, they likely have similar ratios of recent to past averaged SFR as other galaxies with the same luminosity. With a couple of exceptions, the LSB galaxies have $(FUV-NUV) \sim 0.2$ mag, similar to most higher surface brightness galaxies. Stellar population models with  constant star formation history, standard IMF, solar metallicity, and no dust predict colors in the range $(FUV-NUV) \sim 0.0 - 0.2$, with the precise value depending on the particular model. Thus, we do not find strong evidence for variability in the recent star formation histories of our LSB galaxies. 

For a subset of four LSB galaxies with both FIR data from {\it Spitzer} and UV data from {\it GALEX}, we find FIR/UV ratios significantly less than unity, indicating that most of the light from the young stars escapes unimpeded from these galaxies and is not absorbed by dust. While more FIR data for LSB galaxies would be desirable, we assumed that  all of our LSB galaxies have negligible amounts of dust and that the UV fluxes alone can be used to determine their SFRs.

We used the UV images to determine the average SFR surface density and published \ion{H}{1} data to determine the gas surface density. While there is little data on the molecular content of LSB galaxies, the available detections and upper limits are consistent with molecular fractions less than 10\% assuming a standard Galactic CO to H$_2$ conversion factor. Thus we assumed for the LSB galaxies that $\Sigma_{gas} \approx \Sigma_{HI}$. In the first version of the star formation law in which $\Sigma_{SFR}$ is plotted as a function of $\Sigma_{gas}$, the LSB galaxies lie about a factor of five below the 1.4 power law fit to the high surface brightness sample from \citet{kennicutt1998a}.  The azimuthally averaged radial SFR and gas density profiles for the LSB galaxies tend to lie in the same region as the integrated measurements. Given the resolution of the \ion{H}{1} maps there is no indication of local variations in the star formation law except in the centers of two of the giant LSB galaxies, where the UV emission may be contaminated by light from an AGN and where there may be some amount of molecular gas. In the second version of the star formation law in which $\Sigma_{SFR}$ is plotted as a function $\Sigma_{gas} / \tau_{dyn}$, the LSB galaxies similarly lie below the extrapolation of the fit to the higher surface brightness sample allbeit with more scatter than in the star formation law as a function of $\Sigma_{gas}$ alone. 

The downturn observed in both star formation relations at densities below about $10 M_{\odot}$ pc$^{-2}$ would be consistent with a lower mean star formation efficiency in LSB galaxies. The observed down-turn is similar to that predicted theoretically by \citet{krumholz2005}. In their model star formation occurs in supersonically turbulent molecular clouds. This model coupled with the relation between the molecular fraction and ISM pressure derived by \citet{blitz2006} predicts a downturn at $\Sigma_{gas} < 10 M_{\odot}$ pc$^{-2}$ in both forms of the star formation law due to the declining molecular fraction with decreasing gas and stellar density. Indeed a plot of $\Sigma_{SFR}$ as a function of the molecular gas surface density alone shows that on average the LSB galaxies have similar star formation efficiencies to higher surface brightness galaxies when only considering the molecular gas. On the other hand the scatter in the $\Sigma_{SFR} - \Sigma_{H2}$ relation is significantly larger than that in the $\Sigma_{SFR} - \Sigma_{HI + H2}$ relation. If this scatter is real, then there could be some other parameter in addition to  the molecular gas  content that determines the SFRs for galaxies. However, the scatter may be simply due to variations in the CO to H$_2$ conversion factor as a function of metallicity or density. Clearly more data on variations in this conversion factor, as well as better constraints on the molecular gas in LSB galaxies are needed to better test these alternatives.

\acknowledgments
We wish to thank Joannah Hinz, Samuel Boissier, Robert Kennicutt, and Mark Krumholz for helpful discussions. GALEX (Galaxy Evolution Explorer) is a NASA Small Explorer, launched in April 2003. We gratefully acknowledge NASA's support for construction, operation, and science analysis for the GALEX mission, developed in cooperation with the Centre National d'Etudes Spatiales of France and the Korean Ministry of Science and Technology.

\clearpage

\clearpage

\begin{deluxetable}{lccrr}
\tablecaption{LSB Galaxy Sample}
\tablewidth{0pt}
\tablehead{ \colhead{Galaxy} & \colhead{RA (2000)} & \colhead{Dec (2000)} & \colhead{Distance} & \colhead{$E(B-V)$\tablenotemark{a}} \\ & & & (Mpc) & \colhead{mag}}
\startdata
    LSBC F561-01  &    $08^{h}09^{m}41.3^{s}$   & $+22^{\circ}33\arcmin38\arcsec$  &  67  & 0.05 \\
    LSBC F563-01  &    08 55 07.2   & +19 45 01  &   49 & 0.03  \\
   LSBC F563-V01  &    08 46 37.8   & +18 53 21  &    54  & 0.03\\
   LSBC F564-V03  &    09 02 53.9   & +20 04 31  &      9  & 0.03\\
   LSBC F565-V02  &    09 37 30.7   & +21 45 59  &   51  & 0.03\\
    LSBC F568-01  &    10 26 06.4   & +22 26 02  &    91  & 0.02 \\
    LSBC F568-03  &    10 27 20.3   & +22 14 27  &    83  & 0.02 \\
   LSBC F568-V01  &    10 45 02.0   & +22 03 17  &    86  & 0.03 \\
   LSBC F571-V02  &    11 37 29.3   & +18 24 40  &      17  & 0.03 \\
    LSBC F574-01  &    12 38 07.3   & +22 18 50  &    103  & 0.02 \\
    LSBC F574-02  &    12 46 43.5   & +21 49 52  &     94  & 0.04 \\
   LSBC F577-V01  &    13 50 10.1   & +18 16 06  &    114  & 0.02 \\
   LSBC F579-V01  &    14 32 49.9   & +22 45 41  &     91  & 0.03 \\
    LSBC F583-01  &    15 57 27.5   & +20 39 58  &    34 & 0.05 \\
    LSBC F568-06  &    10 39 52.5   & +20 50 49  &   205  & 0.03 \\
   Malin 1  &    12 36 59.4   & +14 19 49  &   380 & 0.04 \\
  UGC 6614  &    11 39 14.9   & +17 08 36  &     92  & 0.03 \\
  UGC 5750  &    10 35 45.1   & +20 59 24  &     60 & 0.02  \\
  UGC 5999  &    10 52 59.1   & +07 37 11  &    49 & 0.03  \\
\enddata
\tablenotetext{a}{Galactic reddening from \citet{schlegel1998}.}
\label{sample}
\end{deluxetable}

\begin{deluxetable}{llrr}
\tablecaption{{\it GALEX} observations}
\tablewidth{0pt}
\tablehead{\colhead{Galaxy} & \colhead{Tilename} & \colhead{FUV} & \colhead{NUV} 
\\ & & (sec) & (sec)}
\startdata
     LSBC F561-01 &                   AIS\_195\_sg19 &   111 &   111 \\
   LSBC F563-01 &                  NGA\_LSBC\_D563 &  1696 &  1696 \\
  LSBC F563-V01 &                   AIS\_193\_sg30 &   110 &   110 \\
  LSBC F564-V03 &                   AIS\_193\_sg57 &   110 &   110 \\
  LSBC F565-V02 &                  NGA\_LSBC\_F565 &  1683 &  1683 \\
   LSBC F568-01 &                   AIS\_333\_sg36\tablenotemark{a} &    87 &   295 \\
   LSBC F568-03 &            GI1\_047040\_UGC05672 &  1772 &  1772 \\
  LSBC F568-V01 &                   AIS\_482\_sg61 &   110 &   199 \\
  LSBC F571-V02 &                   AIS\_226\_sg32 &    91 &    91 \\
   LSBC F574-01 &                   AIS\_222\_sg69 &    95 &    95 \\
   LSBC F574-02 &                   AIS\_222\_sg86 &   104 &   258 \\
  LSBC F577-V01 &                   AIS\_217\_sg69 &    96 &    96 \\
  LSBC F579-V01 &                   AIS\_216\_sg33 &    80 &    87 \\
   LSBC F583-01 &      GI1\_073008\_J155912p204531 &  1572 &  2609 \\
   LSBC F568-06 &                   AIS\_333\_sg70 &    95 &   195 \\
  Malin 1 &                     NGA\_MALIN1 &  1819 &  1819 \\
 UGC 6614 &                    NGA\_UGC6614 &     0 &   679 \\
 UGC 5750 &            GI1\_067010\_UGC05750 &  3731 &  3731 \\
 UGC 5999 &                   AIS\_312\_sg84 &   108 &   217 \\
\enddata
\label{uvobs}
\tablenotetext{a}{The publicly available version of the tile AIS\_333\_sg36 in GR4 only includes data from two $NUV$-only visits. Since the third visit to this tile does include $FUV$ data, we have made our own custom co-add for use in this paper.}
\end{deluxetable}

\begin{deluxetable}{lrrrccc}
\tablecaption{Photometry aperture definitions and sky values}
\rotate
\tablewidth{0pt}
\tablehead{\colhead{Galaxy} & \colhead{Axis Ratio} & \colhead{P.A.} & 
\colhead{$r_{max}$} & \colhead{FUV sky} & \colhead{NUV sky} & \colhead{$r$ sky} \\
 & & \colhead{(deg)} & \colhead{(arcsec)} & \colhead{($10^{-4}$ cts/s/pix)} & \colhead{($10^{-3}$ cts/s/pix)} &\colhead{(ADU/s/pix)}}
\startdata

   LSBC F561-01  & 0.92  &  55  &  33  &  $4.18\pm0.30$  &  $4.29\pm0.33$  &   $ 1.984\pm 0.002$  \\
   LSBC F563-01  & 0.91  & 341  &  81  &  $2.16\pm0.11$  &  $2.99\pm0.12$  &   $ 2.481\pm 0.007$  \\
  LSBC F563-V01  & 0.52  & 320  &  30  &  $2.37\pm0.33$  &  $3.30\pm0.14$  &   $ 2.512\pm 0.009$  \\
  LSBC F564-V03  & 0.87  & 156  &  40  &  $2.57\pm0.35$  &  $3.32\pm0.17$  &   $ 2.395\pm 0.004$  \\
  LSBC F565-V02  & 0.52  & 205  &  36  &  $2.29\pm0.06$  &  $3.05\pm0.04$  &   $ 2.199\pm 0.004$  \\
   LSBC F568-01  & 0.90  &  13  &  36  &  $1.74\pm0.29$  &  $2.68\pm0.10$  &   $ 1.916\pm 0.002$  \\
   LSBC F568-03  & 0.77  & 169  &  44  &  $1.98\pm0.11$  &  $2.68\pm0.05$  &   $ 2.161\pm 0.003$  \\
  LSBC F568-V01  & 0.77  & 316  &  40  &  $1.85\pm0.49$  &  $3.03\pm0.21$  &   $ 2.145\pm 0.021$  \\
  LSBC F571-V02  & 0.72  & 210  &  48  &  $2.42\pm0.31$  &  $2.74\pm0.19$  &   $ 2.665\pm 0.003$  \\
   LSBC F574-01  & 0.44  & 100  &  33  &  $2.78\pm0.42$  &  $2.73\pm0.11$  &   $ 2.336\pm 0.001$  \\
   LSBC F574-02  & 0.87  &  53  &  25  &  $1.96\pm0.37$  &  $2.86\pm0.06$  &   $ 2.293\pm 0.004$  \\
  LSBC F577-V01  & 0.82  &  40  &  22  &  $1.99\pm0.19$  &  $2.38\pm0.11$  &   $ 2.505\pm 0.002$  \\
  LSBC F579-V01  & 0.90  & 120  &  42  &  $2.43\pm0.27$  &  $2.34\pm0.14$  &   $ 1.652\pm 0.002$  \\
   LSBC F583-01  & 0.47  & 355  &  77  &  $3.68\pm0.22$  &  $2.61\pm0.06$  &   $ 1.691\pm 0.004$  \\
   LSBC F568-06  & 0.80  &  75  &  80  &  $1.95\pm0.33$  &  $2.64\pm0.06$  &   $ 2.211\pm 0.002$  \\
  Malin 1  & 0.72  &  22  &  70  &  $2.37\pm0.16$  &  $2.80\pm0.05$  &   $ 2.632\pm 0.009$  \\
 UGC 6614  & 0.83  & 296  & 125  &        \nodata  &  $3.47\pm0.04$  &   $ 2.897\pm 0.004$  \\
 UGC 5750  & 0.47  & 167  &  58  &  $2.01\pm0.06$  &  $2.96\pm0.05$  &   $ 2.047\pm 0.007$  \\
 UGC 5999  & 0.60  & 131  &  65  &  $2.81\pm0.40$  &  $3.13\pm0.07$  &   $ 2.643\pm 0.004$  \\
\enddata
\label{photaper}
\end{deluxetable}

\begin{deluxetable}{lcccc}
\tablecaption{UV photometry}
\tablewidth{0pt}
\tablehead{\colhead{Galaxy} & \colhead{FUV} & \colhead{NUV} & \colhead{$<\mu>_{FUV}$}  & 
\colhead{$<\mu>_{NUV}$} \\
 & \colhead{(AB mag)} & \colhead{(AB mag)} & \colhead{(mag arcsec$^{-2}$)} & 
 \colhead{(mag arcsec$^{-2}$})}
\startdata
   LSBC F561-01  &    $18.17 \pm 0.11$  &    $17.58 \pm 0.09$  &  26.91  &  26.32  \\
 LSBC F563-01  &    $17.50 \pm 0.06$  &    $17.26 \pm 0.10$  &  28.18  &  27.94  \\
LSBC F563-V01  &    $19.67 \pm 0.24$  &    $19.13 \pm 0.13$  &  27.58  &  27.04  \\
LSBC F564-V03  &    $18.98 \pm 0.18$  &    $18.87 \pm 0.18$  &  28.09  &  27.98  \\
LSBC F565-V02  &    $19.29 \pm 0.07$  &    $19.03 \pm 0.06$  &  27.59  &  27.34  \\
 LSBC F568-01  &    $18.30 \pm 0.12$  &    $17.99 \pm 0.07$  &  27.21  &  26.90  \\
 LSBC F568-03  &    $17.86 \pm 0.06$  &    $17.62 \pm 0.06$  &  27.04  &  26.80  \\
LSBC F568-V01  &    $18.19 \pm 0.12$  &    $18.04 \pm 0.09$  &  27.17  &  27.01  \\
LSBC F571-V02  &    $18.15 \pm 0.12$  &    $17.80 \pm 0.10$  &  27.43  &  27.08  \\
 LSBC F574-01  &    $18.39 \pm 0.12$  &    $18.16 \pm 0.08$  &  26.34  &  26.11  \\
 LSBC F574-02  &    $19.03 \pm 0.15$  &    $18.92 \pm 0.08$  &  27.11  &  27.00  \\
LSBC F577-V01  &    $18.31 \pm 0.12$  &    $18.20 \pm 0.08$  &  26.11  &  25.99  \\
LSBC F579-V01  &    $18.40 \pm 0.14$  &    $18.12 \pm 0.11$  &  27.64  &  27.36  \\
 LSBC F583-01  &    $17.65 \pm 0.07$  &    $17.46 \pm 0.06$  &  27.54  &  27.35  \\
 LSBC F568-06  &    $17.52 \pm 0.13$  &    $17.24 \pm 0.07$  &  28.04  &  27.75  \\
     Malin 1  &    $19.25 \pm 0.15$  &    $19.01 \pm 0.12$  &  29.36  &  29.12  \\
    UGC 6614  &             \nodata  &    $16.30 \pm 0.06$  &\nodata  &  27.82  \\
    UGC 5750  &    $17.44 \pm 0.06$  &    $17.29 \pm 0.06$  &  26.72  &  26.57  \\
    UGC 5999  &    $17.03 \pm 0.08$  &    $16.86 \pm 0.06$  &  26.77  &  26.61  \\
\enddata
\tablecomments{All the magnitudes and surface brightnesses in this table are in the AB system and are uncorrected for Galactic extinction. The surface brightness here is defined to be the average surface brightness within the entire elliptical aperture as given in Table \ref{photaper}. The surface brightnesses have not been corrected for inclination or redshift.}
\label{uvphot}
\end{deluxetable}

\begin{deluxetable}{lcrr}
\tablecaption{Optical photometry}
\tablewidth{0pt}
\tablehead{\colhead{Galaxy} & \colhead{$r$} & \colhead{$r_{1/2}$\tablenotemark{a}} & 
\colhead{$\mu_{r,1/2}$} \\
& \colhead{(AB mag)} & \colhead{(arcsec)} & \colhead{(mag arcsec$^{-2}$})}
\startdata
 LSBC F561-01  &    $15.67 \pm 0.02$  &     14  &  23.43  \\
 LSBC F563-01  &    $15.78 \pm 0.20$  &     13  &  23.29  \\
LSBC F563-V01  &    $16.74 \pm 0.12$  &     14  &  23.89  \\
LSBC F564-V03  &    $16.37 \pm 0.15$  &     17  &  24.37  \\
LSBC F565-V02  &    $16.95 \pm 0.10$  &     21  &  24.93  \\
 LSBC F568-01  &    $15.95 \pm 0.04$  &     14  &  23.60  \\
 LSBC F568-03  &    $15.48 \pm 0.04$  &     14  &  22.98  \\
LSBC F568-V01  &    $16.23 \pm 0.26$  &     11  &  23.21  \\
LSBC F571-V02  &    $15.23 \pm 0.04$  &     22  &  23.64  \\
 LSBC F574-01  &    $16.00 \pm 0.01$  &     13  &  22.79  \\
 LSBC F574-02  &    $16.79 \pm 0.06$  &     13  &  24.28  \\
LSBC F577-V01  &    $16.75 \pm 0.03$  &     12  &  23.97  \\
LSBC F579-V01  &    $15.40 \pm 0.03$  &     14  &  23.03  \\
 LSBC F583-01  &    $15.92 \pm 0.16$  &     16  &  23.19  \\
 LSBC F568-06  &    $13.70 \pm 0.02$  &     22  &  22.17  \\
     Malin 1  &    $16.21 \pm 0.19$  &      5  &  21.72  \\
    UGC 6614  &    $13.11 \pm 0.06$  &     21  &  21.52  \\
    UGC 5750  &    $15.66 \pm 0.13$  &     18  &  23.23  \\
    UGC 5999  &    $15.15 \pm 0.06$  &     28  &  23.84  \\
\enddata
\tablenotetext{a}{The half-light radii were determined from the $r$-band radial profile.}
\tablecomments{All the magnitudes and surface brightnesses in this table are in the AB system and are uncorrected for Galactic extinction.}
\label{sdssphot}
\end{deluxetable}

\begin{deluxetable}{lccccc}
\tablecaption{Star Formation Law Data}
\tablewidth{0pt}
\tablehead{\colhead{Galaxy} & \colhead{log SFR} & \colhead{log $\Sigma_{SFR}$} & 
\colhead{log $\Sigma_{HI}$} & \colhead{$V_{rot}$} & \colhead{$\tau_{dyn}$} \\
 & \colhead{($M_{\odot}$ yr$^{-1}$)} & \colhead{($M_{\odot}$ yr$^{-1}$ kpc$^{-2}$)} & 
 \colhead{($M_{\odot}$ pc$^{-2}$)} & \colhead{(km s$^{-1}$)} & \colhead{($10^8$ yr)}
 }
\startdata
LSBC F561-01 & -0.68 & -3.21 & 0.62 &  50 &    9 \\
LSBC F563-01 & -0.76 & -3.79 & 0.57 & 111 &    7 \\
LSBC F563-V01 & -1.53 & -3.80 & 0.51 &  29 &   11 \\
LSBC F564-V03 & -2.91 & -3.79 & 0.32 & \nodata &    \nodata \\
LSBC F565-V02 & -1.42 & -3.80 & 0.39 &  43 &    8 \\
LSBC F568-01 & -0.55 & -3.42 & 0.86 & 119 &    5 \\
LSBC F568-03 & -0.46 & -3.41 & 0.63 & 108 &    7 \\
LSBC F568-V01 & -0.54 & -3.44 & 0.58 & 111 &    6 \\
LSBC F571-V02 & -1.92 & -3.60 & 0.52 &  42 &    4 \\
LSBC F574-01 & -0.48 & -3.36 & 0.44 &  96 &    7 \\
LSBC F574-02 & -0.75 & -3.33 & 0.49 &  29 &   16 \\
LSBC F577-V01 & -0.36 & -3.00 & 0.37 &  30 &   17 \\
LSBC F579-V01 & -0.57 & -3.56 & 0.50 & 117 &    6 \\
LSBC F583-01 & -1.06 & -3.76 & 0.63 &  85 &    6 \\
LSBC F568-06 &  0.48 & -3.74 & 0.28 & 286 &   12 \\
Malin 1 &  0.36 & -4.21 & 0.35 & 187 &   29 \\
UGC 6614 &  0.29 & -3.67 & 0.42 & 214 &   11 \\
UGC 5750 & -0.57 & -3.51 & 0.43 &  70 &   10 \\
UGC 5999 & -0.54 & -3.39 & 0.67 & 140 &    4 \\
\enddata
\tablecomments{The values for $\Sigma_{SFR}$ were calculated from the UV surface brightnesses in Table \ref{uvphot} assuming a \citet{salpeter1955} stellar IMF after correcting for Galactic extinction, redshift surface brightness  dimming, and inclination. The values for $\Sigma_{gas}$, $V_{rot}$, and $\tau_{dyn}$ are derived from \ion{H}{1} observations \citep{vanderhulst1993, deblok1996, pickering1997}.}
\label{sflaw_table}
\end{deluxetable}

\begin{deluxetable}{lccc}
\tablecaption{Molecular Star Formation Law Data}
\tablewidth{0pt}
\tablehead{\colhead{Galaxy} & \colhead{$\Sigma_{H2}$} & \colhead{$\Sigma_{SFR}$} & 
\colhead{CO ref.} \\
& \colhead{($M_{\odot}$ pc$^{-2}$)} & \colhead{($10^{-4}$ $M_{\odot}$ yr$^{-1}$ kpc$^{-2}$)}
}
\startdata
LSBC F561-01 &  $<1.8$ & 8.8 & 1 \\
LSBC F563-01 &  $<1.6$ & 8.3 & 1 \\
LSBC F571-V02 &  $<0.9$ & 6.4 & 1 \\
LSBC F583-01 &  $<2.2$ & 8.7 & 1 \\
LSBC F563-V01 &  $<1.3$ & 6.6 & 2 \\
LSBC F568-V01 &  $<1.4$ & 17 & 2 \\
Malin 1 &  $<0.7$ & 0.6 & 3 \\
UGC 6614 &  $0.2$ & 28 & 4 \\
\enddata
\tablecomments{The molecular gas surface densities and upper limits assume a conversion factor between CO brightness temperature and molecular gas surface density of 4.5 $M_{\odot}$ pc$^{-2}$ (K km s$^{-1}$)$^{-1}$. The UV surface brightnesses were measured in the same aperture as used for the CO observations.}
\tablerefs{(1) \citet{schombert1990}; (2) \citet{deblok1998b}; (3) \citet{braine2000}; (4) \citet{das2006}}
\label{sflaw_mol_table}
\end{deluxetable}

\end{document}